\documentclass[aps,reprint,groupedaddress,longbibliography]{revtex4-1}
\pdfpageattr {/Group << /S /Transparency /I true /CS /DeviceRGB>>}
\usepackage{amsmath}
\usepackage{amsthm}
\usepackage{physics}
\usepackage{amsfonts}
\usepackage{graphicx}
\usepackage{wrapfig}
\usepackage{times,txfonts}
\usepackage[colorlinks=true,linkcolor=blue,urlcolor=blue,citecolor=blue,pdfusetitle]{hyperref}
\usepackage{hyperref}
\usepackage[dvipsnames]{xcolor}

\renewcommand{\ket}[1]{\ensuremath{|#1\rangle}}
\renewcommand{\bra}[1]{\langle#1|}
\renewcommand{\braket}[2]{\langle#1|#2\rangle}

\begin{document}
	\title{Thermometry of Strongly Correlated Fermionic Quantum Systems using Impurity Probes}
	
	\author{George Mihailescu}
	\email[]{george.mihailescu@ucdconnect.ie}
	\affiliation{School of Physics, University College Dublin, Belfield, Dublin 4, Ireland}
	\affiliation{Centre for Quantum Engineering, Science, and Technology, University College Dublin,  Dublin 4, Ireland}
	\author{Steve Campbell}
	\email[]{steve.campbell@ucd.ie }
	\affiliation{School of Physics, University College Dublin, Belfield, Dublin 4, Ireland}
	\affiliation{Centre for Quantum Engineering, Science, and Technology, University College Dublin, Dublin 4, Ireland}
	\author{Andrew K. Mitchell}
	\email[]{andrew.mitchell@ucd.ie}
	\affiliation{School of Physics, University College Dublin, Belfield, Dublin 4, Ireland}
	\affiliation{Centre for Quantum Engineering, Science, and Technology, University College Dublin,  Dublin 4, Ireland}


\begin{abstract}
We study quantum impurity models as a platform for quantum thermometry. A single quantum spin-$\tfrac{1}{2}$ impurity is coupled to an explicit, structured, fermionic thermal environment which we refer to as the environment or bath. We critically assess the thermometric capabilities of the impurity as a probe, when its coupling to the environment is of Ising or Kondo exchange type. In the Ising case, we find sensitivity equivalent to that of an idealized two-level system, with peak thermometric performance obtained at a temperature that scales linearly in the applied control field, independent of the coupling strength and environment spectral features. By contrast, a richer thermometric response can be realized for Kondo impurities, since strong probe-environment entanglement can then develop. 
At low temperatures, we uncover a regime with a universal thermometric response that is independent of microscopic details, controlled only by the low-energy spectral features of the environment. The many-body entanglement that develops in this regime means that low-temperature thermometry with a weakly applied control field is inherently less sensitive, while optimal sensitivity is recovered by suppressing the entanglement with stronger fields.
\end{abstract}
\maketitle


\section{Introduction}
The characterization of a system through its physical and thermodynamic observables is not only of foundational importance, but crucial in the design, control, and manipulation of experimental and commercial platforms. Moreover, it is vital that one has access to high-precision measurements of the operating parameters of devices in the NISQ era, where mitigating the deleterious effects of noise is a central task. In particular, having precise knowledge of the temperature is required to assess whether a given system is in the correct operating regime. 
However, thermometry for quantum systems is challenging in part since there is no unique observable one can assign to temperature~\cite{Temp_in_out_Equil,Col_Quant_Fluc_Rel,Char_Work_Fluc,Entropy_Prod_Cor_sys_res,binder2018thermodynamics}. Therefore, we must resort to quantum estimation techniques which provide strategies for inferring unknown parameters and place fundamental bounds on how accurately we may measure them~\cite{Paris_Quantum_Estimation, Giovannetti_Quantum_Metrology, helstrom1969quantum}. The importance of temperature estimation, particularly for low temperature systems~\cite{CorreaPRB}, has led to a plethora of sensing platforms such as minimal thermometers comprised of individual quantum probes~\cite{Campbell_Percision_Thermo_QSL} and more complicated spin systems~\cite{Mehboudi_Thermo_Per_Strog_Corr_Lattice}. Various thermometric strategies have been suggested~\cite{Mehboudi_Thermo_Review}, for example exploiting the non-equilibrium dynamics of the probe~\cite{MitchisonPRL, Cavina_Thermo_Metrology_NonEquil_Thermo,Razavian2019} to infer the temperature, collisional schemes~\cite{OConnor_Col,Surpassing_CR_Col_Ther,Landi_Col,Q_Col_thermo,baris_2020}, global measurement schemes~\cite{bayat_optimal_probs},  or leveraging quantum criticality to endow thermometers with enhanced sensitivity~\cite{Puebla_Exp_Per_QCP, Puebla_Quantum_Metrology_Fully_Connected_Models, Paris_LMG_Critical_Systm,Tommaso_Quantum_Critical_Metrology,Wald_Quantum_Metrology_MF_Criticality,zhang2022non}. 

In this work we address an important aspect of quantum sensing: the role that the microscopic description of the sample plays in the achievable thermometric precision~\cite{ClaudiaSpec1, ClaudiaSpec2}, and how the development of quantum entanglement between the probe and sample affects this precision. Our setting consists of an explicit probe-sample composite that we assume have reached thermal equilibrium with one another at some ambient temperature. We regard the sample as a quantum environment or `bath' whose temperature the probe is designed to estimate. We focus here on fermionic (electronic) environments in the thermodynamic limit. The probe will be a single spin-$\tfrac{1}{2}$ quantum `impurity' (qubit) embedded in the sample. We examine how the nature of the probe-environment coupling, as well as the spectral properties of the environment itself, determine the sensing capabilities of the probe. We compare two paradigmatic quantum impurity models to describe the coupled probe-environment system. 

First, we consider an Ising-like coupling, which allows for an exact analytic solution, even for baths with an infinite number of degrees of freedom, and with arbitrary spectral properties. We find that the thermometric sensitivity of the probe in this setting is essentially independent of the environment's spectral features and, since this coupling does not give rise to any entanglement between the probe and the environment, the probe behaves effectively like a free spin in an \emph{implicit} thermal environment~\cite{correa2015individual,jevtic2015single}. 

Secondly, we allow for spin-flip (exchange) terms in the interaction Hamiltonian, thus realising the so-called impurity Kondo model~\cite{kondo1964resistance}. This coupling gives rise to a significantly richer behavior as the probe and bath become entangled. In this case we find nontrivial temperature and control-field dependences of the metrological precision of the impurity probe, and observe that the bath's spectral features now play a crucial role.
We uncover a characteristic universality in the  sensitivity of this probe at low temperatures, indicating that the same thermometric response is obtained independently of bare system parameters and can be used `model free'.


\section{Preliminaries}
\subsection{Quantum Thermometry for Equilibrium Systems}
A typical thermometric protocol involves allowing a probe system, the thermometer, to reach thermal equilibrium with the sample~\cite{Paris_Quantum_Estimation,Mehboudi_Thermo_Review}, which hereafter we refer to as the environment or bath. Ideally, the probe acts as a weak perturbation to the environment and acts non-invasively~\cite{CampbellNJP2017,Probe_Inac_Envo,Ortho_Cath_Qubit_Fermi,Prob_Hyb_Parial_Ac,Nondes_prob_Bose}. However, for quantum systems, correlations and entanglement can build up between the probe and the environment, especially at low temperatures, with the nature of the probe-environment entanglement often delicately dependent on the manner and strength of their mutual interaction and the spectral properties of the environment~\cite{Quantum_Limits_Thermo,Correa_Low_T_Strong_Cop,Mehboudi_Thermo_Per_Strog_Corr_Lattice,Mehboudi_Bath_Induced_Corr_Low_Tem,Paris_Tight_Bound_Thermo_Low_T,Brunner_Fundamental_Limit_Low_T_Quantum__Thermo,de2016local}. The thermometric precision attainable in such an equilibrium setting will be the focus of this work. Specifically, we consider explicit fermionic environments in the thermodynamic limit, probed by spin-qubit `impurities'. 

A given probe will be deemed a good thermometer if it has a high sensitivity over a tunable target temperature range, and if it is `robust' in the sense that it works in different environment settings without the need for detailed information about the structure of that environment. 

The framework of equilibrium temperature sensing considers an arbitrary $n$-dimensional system with Hamiltonian $\hat{H}\! =\! \sum_n E_n\ket{E_n}\bra{E_n}$, where the system is described by the Gibbs state $\hat{\varrho}_T\left(T\right) \!=\!  \tfrac{1}{Z}e^{-\hat{H}/T}\! =\! \sum_n p_n\ket{E_n}\bra{E_n}$, with  $\{\ket{E_n}\}$ the Hamiltonian eigenstates and $\{p_n\}$ their thermal populations, with partition function $Z \!=\! {\rm Tr}\left[{e^{-\hat{H}/T}}\right]$ \cite{Mehboudi_Thermo_Review,correa2015individual}. We employ tools from quantum parameter estimation theory to assess the thermometric precision \cite{helstrom1969quantum,Paris_Quantum_Estimation}. Within this framework, we can estimate some unknown quantity $\lambda$, whose information is imprinted on a quantum state, $\varrho(\lambda)$, by performing a measurement of an arbitrary observable. Repeating this process many times and building up a dataset of outcomes, we build an estimator $\bar{\lambda}$. Statistical errors arising from the uncertainty of these outcomes are unavoidable, and the precision with which the parameter of interest can be estimated is quantified through the Cram\'er-Rao bound~\cite{CRB_Book}, 
\begin{equation}
	\label{Eq:Thermometry_CRB}
	\text{Var}\left(\bar{\lambda}\right) \geq \frac{1}{{\mathcal{M}}{F\left(\lambda\right)}} \;,
\end{equation}
which sets a lower-bound on the statistical error via the variance of the parameter, with $\mathcal{M}$ the number of measurements performed and $F(\lambda)$ the Fisher information (FI) associated to the particular choice of measurement. The FI quantifies the amount of information that a given observable carries about the unknown parameter $\lambda$,
\begin{equation}
	\label{Eq:Thermometry_Generic_FI}
	F\left(\lambda\right) = \sum_k{\abs{{\partial_\lambda}{\ln{p_k}}}^2}{p_k} \;,
\end{equation}
for a discrete set of outcomes with probability, $p_k$, of obtaining outcome $k$ from the measurement. In the context of parameter estimation theory, the FI gives the sensitivity with which we can estimate the parameter $\lambda$ -- where evidently, this sensitivity is intimately related to the choice of measurement employed. The quantum Fisher information (QFI) allows us to determine the maximal attainable sensitivity by maximising the FI overall possible measurements, and is given by,
\begin{equation}
	\label{Eq:Thermometry_Generic_QFI}
	\mathcal{H}\left(\lambda\right) = \sum_k\frac{\left[{\partial_\lambda}{\varrho_k\left(\lambda\right)}\right]^2}{\varrho_k\left(\lambda\right)} + {2}\sum_{m \neq n} \sigma_{mn} \abs{\braket{\psi_m}{\partial_\lambda \psi_n}}^2 \;,
\end{equation}
where the first term represents the classical FI, with $\varrho_k$ and $\ket{\psi_j}$ the parameter-dependant density matrix eigenvalues and eigenvectors respectively, and the second term captures the quantum contribution to the FI, with $\sigma_{mn}\!= \frac{\left(\varrho_n - \varrho_m\right)^2}{\varrho_n + \varrho_m}$~\cite{QFI_Ref1,QFI_Ref2,Qdv_Quan_Metr,Paris_Quantum_Estimation}. We can apply this framework to quantum thermometry by setting $\lambda \!=\! T$. For a family of Gibbs states $\hat{\varrho}_T\left(T\right)$, the QFI is equivalent to the classical Fisher information corresponding to an energy measurement on the eigenstates of the Hamiltonian $\hat{H}$~\cite{Mehboudi_Thermo_Review}. We use the quantum signal to noise ratio (QSNR),
\begin{equation}
	\label{QSNR}
	\mathcal{Q} = {T}{\sqrt{\mathcal{H}\left(T\right)}} \;.
\end{equation}
to characterise the probe's temperature sensitivity~\cite{MitchisonPRL}.


\subsection{Physical system and model}
We study spin-qubit probes for structured many-body fermionic environments. The models considered are in the class of quantum impurity models, which consist of a few strongly interacting quantum degrees of freedom, coupled to a continuum bath of non-interacting conduction electrons. Here the probe is taken to be a spin-$\tfrac{1}{2}$ impurity coupled to an explicit electronic host system, with an arbitrary structure characterized by its electronic density of states (DoS), $\rho(\omega)$.

Although our focus here is on fundamental aspects of thermometry, we note that the use of impurities as \emph{in-situ} probes for the electronic structure of a host material has a long history in the field of condensed matter physics -- both experimental and theoretical. In its original and simplest context, electronic scattering from actual magnetic impurities (such as iron) embedded in bulk metals (such as gold) were studied using resistivity measurements \cite{de1934electrical,kondo1964resistance,costi2009kondo,Hewson}. Since then, impurities have been widely used as spectroscopic probes \cite{ternes2008spectroscopic,hoffman2002imaging,derry2015quasiparticle,*mitchell2015multiple,mitchell2016signatures}. Indeed impurity spins have also been used as nonlocal sensors of magnetism \cite{yan2017nonlocally}.

Quantum impurity models generically take the form,
\begin{equation}
	\label{Eq:Generic_Impurity_Ham}
	\hat{H} = \hat{H}_{bath} + \hat{H}_{imp} + \hat{H}_{imp-bath} \;,
\end{equation}
where $\hat{H}_{bath}$ describes a continuum bath of non-interacting conduction electrons in the thermodynamic limit, $\hat{H}_{imp}$ is the impurity Hamiltonian, and $\hat{H}_{imp-bath}$ describes the coupling between the impurity and the bath. In its diagonal representation, we write the bath Hamiltonian in second quantized form,
\begin{equation}\label{eq:Hbath}
\hat{H}_{bath} = \sum_{k,\sigma}\epsilon_k^{\phantom{\dagger}} \hat{c}_{k\sigma}^{\dagger}\hat{c}_{k\sigma}^{\phantom{\dagger}}  \;,  
\end{equation}
where $\hat{c}_{k\sigma}^{(\dagger)}$ annihilates (creates) an electron in the single-particle momentum state $k$ with spin $\sigma=\uparrow$ or $\downarrow$. Here $\epsilon_k$ is the electronic dispersion, which is determined by the specific structure of the environment being considered.

In the following we consider a single quantum impurity coupled to such a fermionic bath, with total Hamiltonian,
\begin{equation}
	\label{Eq:System_Hamiltonian}
	\hat{H} = \hat{H}_{bath} + {J^z}{\hat{S}_I^z}{\hat{S}_0^z} + \tfrac{1}{2}{J^\perp} \left ( {\hat{S}_I^+}{\hat{S}_0^-} +{\hat{S}_I^-}{\hat{S}_0^+} \right ) + B_I^{\phantom{\dagger}} \hat{S}_I^z + B_0^{\phantom{\dagger}}\hat{S}_0^z \;,
\end{equation}
where $\hat{S}_I^z$, $\hat{S}_I^+$ and $\hat{S}_I^-$ are spin-$\tfrac{1}{2}$ operators for the impurity, while $\hat{S}_0^z=\tfrac{1}{2}\left(\hat{c}_{0\uparrow }^{\dagger}\hat{c}_{0\uparrow }^{\phantom{\dagger}}-\hat{c}_{0\downarrow }^{\dagger}\hat{c}_{0\downarrow }^{\phantom{\dagger}} \right )$, $\hat{S}_0^+=\hat{c}_{0\uparrow }^{\dagger}\hat{c}_{0\downarrow }^{\phantom{\dagger}}$ and $\hat{S}_0^-=\hat{c}_{0\downarrow }^{\dagger}\hat{c}_{0\uparrow }^{\phantom{\dagger}}$ act on the bath electrons at the impurity position. The impurity is taken to be at the origin, and couples locally in real-space to the bath site $\hat{c}_{0\sigma}=\sum_k \xi_k \hat{c}_{k\sigma}$, where $\xi_k$ is the weight of bath eigenstate $k$ at the impurity position. $J^z$ and $J^{\perp}$ are the Ising and spin-flip coupling interactions, respectively, between the impurity spin and the conduction electron spin density at the impurity position, while $B_I$ and $B_0$ are Zeeman magnetic fields acting locally on the impurity and bath, and we fix $B_I=B_0\equiv B$.  Although we have considered here the case of a local magnetic field, we note that essentially identical results are obtained in the case of a globally applied field, since the dominant contribution comes from the local term included in Eq.~\eqref{Eq:System_Hamiltonian}.

Since the impurity-bath coupling is local, the impurity sees the local conduction electron DoS at the origin, $\rho(\omega)$. It is related to the retarded, real-frequency bath Green's function via $\rho(\omega)=-\tfrac{1}{\pi}{\rm Im}~G_{00}^0(\omega)$, where in Zubarev notation  \cite{zubarev1960double} $G_{00}^0(\omega)=\langle\langle \hat{c}_{0\sigma}^{\phantom{\dagger}} ; \hat{c}_{0\sigma}^{\dagger}\rangle\rangle^0_{\omega}$~denotes the Fourier transform of the real-time propagator  $G_{00}^0(t) = -i\theta(t)\langle \{ \hat{c}_{0\sigma}^{\phantom{\dagger}}(t) ,\hat{c}_{0\sigma}^{\dagger}(0)\} \rangle^0$.

An advantage of this formalism is the flexibility one has in choosing the structure of the environment being probed, through the spectral properties of the bath corresponding to different physical systems. We leverage this to assess the thermometric capabilities of an impurity probe embedded within different host environments. Details of the different bath systems used in the following and their corresponding DoS are given in Appendix~\ref{app:bath}.

In the context of quantum nanoelectronics, we note that the Hamiltonian Eq.~\eqref{Eq:System_Hamiltonian} describes real quantum dot devices, for which the impurity magnetization can be experimentally measured in response to an applied field \cite{piquard2023observation}.

We consider two distinct scenarios for the impurity-bath coupling below: {\it (i)} the `Ising impurity' in which $J^z>0$ but $J^{\perp}=0$; and {\it (ii)} the `Kondo impurity' in which both $J^z$ and $J^{\perp}$ are finite. 
In many real systems, the impurity spin results from a localized, half-filled quantum orbital with strong Coulomb repulsion. The Kondo model can be derived as a low-energy effective model from this more microscopic Anderson impurity model \cite{anderson1961localized,Hewson}, and produces a spin-isotropic, antiferromagnetic exchange interaction, $J^z=J^{\perp}\equiv J>0$. This is the situation we consider for the Kondo model in the following. In both Ising and Kondo cases, there is a competition between the impurity-bath coupling $J$, which favors antiparallel alignment of the spins, and the field $B$, which tends to polarize the spins. However the possibility of spin-flip scattering in the Kondo model generates many-body quantum entanglement between the impurity probe and the bath, which is entirely absent in the Ising case, as reviewed further below.


\subsection{Kondo effect}
The Kondo model is a famous paradigm in the theory of strongly correlated electron systems \cite{Hewson}. Even though the bath consists of non-interacting electrons, the presence of an interacting impurity can drive the system to a strong coupling ground state. For a metallic system, the quantum impurity spin is dynamically screened by conduction electrons at low temperatures $T\ll T_K$, where $T_K$ is an emergent low-energy scale known as the Kondo temperature
\begin{equation}
   T_K\sim D e^{-1/\rho_0 J} \;,
\end{equation}
where $T_K$ depends non-perturbatively on the impurity-environment coupling $J$, $\rho_0$ is the bath DoS at the Fermi energy, and $D$ the conduction electron bandwidth. 
Physical observables exhibit universal scaling in terms of the Kondo temperature $T_K$ provided $T_K \ll D$ \cite{Hewson}. In the following, we specify parameters as dimensionless ratios involving either the conduction electron bandwidth $D$ (which is typically of order eV) or the Kondo temperature $T_K$ (which depends exponentially on the coupling $J$ and in practice can range from 1mK to 100K depending on microscopic details of the setup \cite{Hewson,iftikhar2018tunable,piquard2023observation}). 

At low temperatures the impurity forms a macroscopic spin-singlet state and becomes maximally entangled with a large number of conduction electrons in a surrounding `Kondo cloud'~\cite{mitchell2011real,v2020observation, bayat2010negativity,lee2015macroscopic,shim2018numerical,kim2021universal}.
Regarding the impurity as a metrological probe, one might expect that the development of this strong probe-environment entanglement could suppress sensitivity: only a small amount of information about the bath is imprinted locally on the impurity since, due to the entanglement, the reduced state of the impurity is maximally mixed. On the other hand, the Kondo cloud `evaporates' at higher temperatures \cite{mitchell2011real}, and the Kondo effect is destroyed by strong magnetic fields \cite{costi2000kondo} which tend to polarize the impurity and suppress the crucial spin-flip scattering processes. Therefore we expect a subtle interplay of energy scales $T,B,J$ and $T_K$ in understanding the thermometic sensitivity of Kondo probes.

The Kondo effect is a non-perturbative quantum many-body phenomenon, involving a large number of bath degrees of freedom, characterized by strong entanglement, and involving a low-energy emergent scale $T_K$. There is no analytic solution for generic quantum impurity models, and as such, sophisticated numerical techniques must be used. In this work, we use Wilson's numerical renormalization group \cite{wilson1975renormalization,bulla2008numerical} (NRG) method, described in Appendix~\ref{app:kondo}, which provides access to numerically-exact thermodynamical observables at essentially any temperature.

We note that the metallic Kondo effect has been observed in many systems, ranging from actual magnetic impurities in metals \cite{kondo1964resistance,Hewson,costi2009kondo}, semiconductor quantum dot devices \cite{goldhaber1998kondo,iftikhar2018tunable,piquard2023observation}, and single-molecule transistors \cite{liang2002kondo,mitchell2017kondo} among others. More recently it has also been studied in the context of other, non-metallic host systems \cite{fritz2013physics,gonzalez1998renormalization,mitchell2013kondodiv,*mitchell2013quantum,galpin2008anderson,pasupathy2004kondo,mitchell2013kondo,*mitchell2015kondo,polkovnikov2001impurity}. The Kondo model is therefore a realistic model to describe the low-energy physics of quantum spin probes coupled to fermionic environments. 

The Kondo effect does not arise in the Ising limit $J^{\perp}=0$. Although the impurity and bath electrons are correlated, there is no quantum entanglement between the two in this case.


\subsection{Entanglement negativity}
The entanglement between the impurity and bath is important for understanding the probe's thermometric potential. In the context of quantum impurity models, we quantify this using the negativity~\cite{bayat2010negativity, BayatPRLNeg, lee2015macroscopic, shim2018numerical, kim2021universal} which is a well-established entanglement monotone. In general, the negativity between coupled subsystems $A$ and $B$ is defined as, 
\begin{equation}
\label{negativityEq}
\mathcal{N}_{A;B} = \tfrac{1}{2}\left(||\varrho^{T_A}|| -1\right) \;,
\end{equation}
where $\varrho$ is the full density matrix, $||\cdot ||$ denotes a trace norm, and $T_A$ is the partial trace on subsystem $A$.

In the following we consider two negativity measures for the Kondo model: {\it (i)} the negativity between the impurity and the local bath orbital to which it is coupled, denoted $\mathcal{N}_{I;0}$ and {\it (ii)} the negativity between the impurity and the entire bath, denoted $\mathcal{N}_{I;bath}$. In the former case, we use the NRG reduced density matrix \cite{weichselbaum2007sum} of the impurity and local bath site (having traced out the rest of the bath), and compute the partial trace on the impurity to obtain $\mathcal{N}_{I;0}$. The inclusion of the full bath in latter case makes computing the negativity significantly more involved~\cite{shim2018numerical}. However, in the low temperature limit $T\ll T_K$, we may use the analytical expression derived very recently in Ref.~\cite{kim2021universal} which relates $\mathcal{N}_{I;bath}$ to the impurity magnetization,
\begin{equation}
    \label{Eq:Neg_Full_Bath}
    \mathcal{N}_{I;bath} = \frac{1}{2}\sqrt{1-4\expval{\hat{S}^z_I}^2} \qquad : T\ll T_K
\end{equation}
As such, we see that in the low-temperature regime the global impurity-bath entanglement will decrease with increasing magnetic field, as the impurity becomes polarized. Finally, it is important to remark that in our considered settings the negativity is a sufficient but not necessary condition for entanglement. The possible ``sudden death'' of negativity in the Kondo model at finite temperatures~\cite{shim2018numerical} does \emph{not} imply that the impurity is completely disentangled with the bath. 


\section{Impurity Thermometry}
We begin by considering the temperature estimation of a metallic environment via a spin-$\tfrac{1}{2}$ impurity probe. We take the simplest case of a fermionic bath with $\rho(\omega)=\rho_0 \theta(D-|\omega|)$ corresponding to a constant DoS, $\rho_0$, inside a band of halfwidth, $D$. Here we assume that the environment and probe have already reached equilibrium. In this case the probe state is diagonal in its energy eigenbasis and the maximal sensitivity is given by the thermal FI corresponding to an energy measurement~\cite{correa2015individual,Paris_Landau_Percision_Thermo,de2016local,Quantum_Crit_res}. Since the impurity is a two-level system with thermal populations $p_{\uparrow/\downarrow}=\tfrac{1}{2}\pm \expval{\hat{S}_I^z}$  determined completely by the impurity magnetization $\expval{\hat{S}_I^z}$, we can express the QFI as,  
\begin{equation}
	\label{Eq:QFI_Impurity_Model}
	\mathcal{H}\left(T\right) = \frac{\abs{{\partial_T}\expval{\hat{S}_I^z}}^2}{\frac{1}{4} - \expval{\hat{S}_I^z}^2} \;.
\end{equation}
As the temperature sensitivity of the probe is intrinsically related to its magnetization, the latter is a key quantifier in understanding the underlying thermometric capabilities of the probe -- with high sensitivities achieved when the impurity magnetization changes rapidly with temperature. 
We emphasize that the probe magnetization is a natural and experimentally feasible observable. For example, the impurity magnetization was recently measured experimentally as a function of temperature in the Kondo regime of a quantum dot device in Ref.~\cite{piquard2023observation}.

An alternative view is obtained by exploiting a Maxwell relation connecting the temperature derivative of the magnetization $\expval{\hat{S}_I^z}$ appearing in Eq.~\eqref{Eq:QFI_Impurity_Model} to the field dependence of the thermodynamic entropy $S$. Since $\expval{\hat{S}_I^z}=\partial_{B_I} F$ and $S=-\partial_T F$ in terms of the free energy $F$, we can write $\partial_T \partial_{B_I} F = \partial_{B_I} \partial_T F$ and hence that $\partial_T \expval{\hat{S}_I^z} = -\partial_{B_I} S$. Thus, we see that high temperature sensitivity of a spin-qubit probe is achieved when the \emph{entropy} changes rapidly as the \emph{field} is varied. These are complementary and intuitive perspectives.


\subsection{Free spin limit}
We first briefly recapitulate the limit of a free (decoupled) impurity spin. Here we imagine that the probe has reached thermal equilibrium with the environment, but is then adiabatically disconnected from it before a measurement is taken. While this `implicit environment' setup is somewhat idealized, it provides a useful reference point for the proceeding sections where structured environments are explicitly considered. 

The free spin Hamiltonian is taken to be simply $\hat{H}_{FS}=B\hat{S}^z_I$, yielding a partition function $Z=2\cosh(B/2T)$ and magnetization $\expval{\hat{S}_I^z}=-\tfrac{1}{2}\tanh(B/2T)$. From this we obtain the \emph{ideal} thermometric QSNR,
\begin{equation}\label{eq:freespin}
    \mathcal{Q} = \left(\frac{B}{2T}\right ) \sech \left ( \frac{B}{2T}\right ) \qquad : \text{free spin}
\end{equation}
which is a function of the single reduced parameter $x=B/T$. We see that $\mathcal{Q}(x)$ is non-monotonic, vanishing at both small and large $x$. The maximum sensitivity is achieved at $B\simeq 2.4 T$, for which $\mathcal{Q}_{max}\simeq \tfrac{2}{3}$. The thermometric capacity of two-level probes are such that the peak sensitivity scales linearly with applied field $B$~\cite{Mehboudi_Thermo_Review,Campbell_Percision_Thermo_QSL}.


\subsection{Ising Coupling}
We now turn to the Ising impurity case, in which a quantum spin-$\tfrac{1}{2}$ impurity probe is coupled by an Ising interaction to an explicit fermionic bath in the thermodynamic limit, i.e. Eq.~\eqref{Eq:System_Hamiltonian} with $J^{\perp}=0$. This extends on previous studies of minimal thermometers that assume an implicit, structureless environment, characterized by a canonical Gibbs ensemble~\cite{Mehboudi_Thermo_Review,correa2015individual,de2017estimating}. 

\begin{figure}[t]
	\centering
	\includegraphics[width=1.04\linewidth]{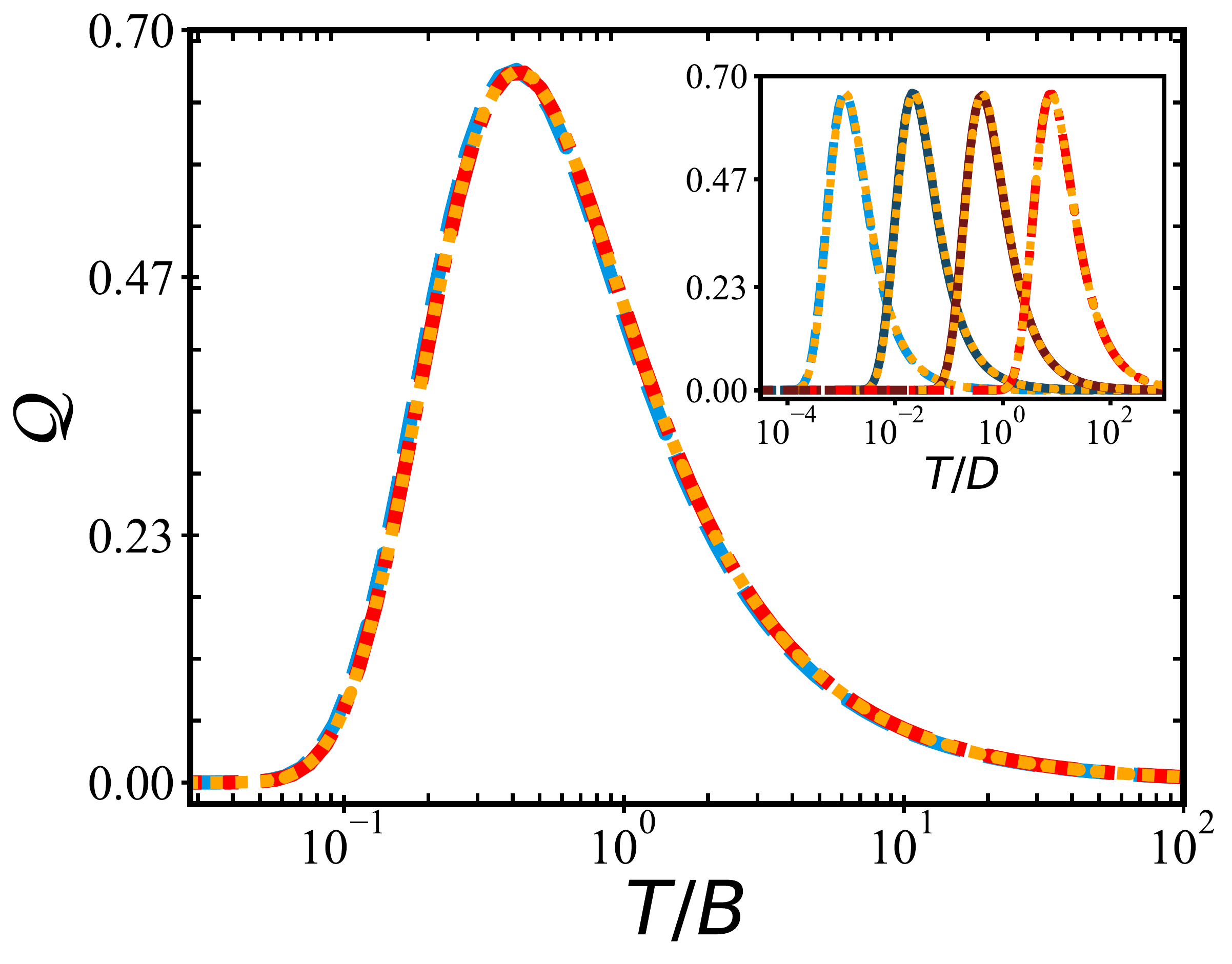}
	\caption{\textbf{\textit{Main Panel:}} Quantum signal to noise ratio (QSNR), $\mathcal{Q}$, for the flat band (metallic) Ising impurity model, plotted as a function of $T/B$. We fix $J^z\! =\! 0.1D$ , varying $T$ for $B/D \!=\! 10^{-2}$ (dashed blue line) and $10^{1}$ (dashed red), demonstrating universal scaling collapse to the free spin (Gibbs state) result, Eq.~\eqref{eq:freespin} (dotted orange). \textbf{\textit{Inset:}} $\mathcal{Q}$ vs $T/D$ for $B/D=10^{-2}, 10^{-1}, 10^0, 10^1$ going from left to right. Again here we also show the free spin result in each case (dotted lines).} 
\label{Fig:FB_Ising_Impurity_FS_Multiple}
\end{figure}

We note that since $[\hat{H},\hat{S}^z_I]=0$ for the Ising impurity, the impurity spin projection is a conserved quantity. This means that there is in fact no equilibration mechanism through which the impurity can thermalize. Nevertheless, we assume here that the impurity and bath are at thermal equilibrium. In principle one could achieve this by initially preparing the system with a finite $J^\perp$ which is then adiabatically turned off, or by starting out with an infinitesimal $J^\perp$ in the distant past. The conserved impurity spin projection in the Ising impurity limit means that the model can be solved analytically for any arbitrary electronic bath, as shown explicitly in Appendix~\ref{app:ising}. Remarkably, our exact result shows that in the case of an infinitely wide flat conduction band, or for general baths but with $B_0=0$, the impurity magnetization is \textit{precisely} that of the free spin. This means that in these special cases, the probe sensitivity is given by Eq.~\eqref{eq:freespin}. 

For the relevant case of finite $B_0=B_I\equiv B$ and finite conduction electron bandwidth $D$, we expect corrections to the free spin (implicit bath) result, as detailed in Appendix~\ref{app:ising}. However, in practice we find these corrections to be very small, except at very large $J^z \sim D$.  Fig.~\ref{Fig:FB_Ising_Impurity_FS_Multiple} shows the thermometric QSNR, plotted for a flat band Ising impurity model using fixed $J_z=0.1D$ and different $T$ and $B$. In the inset, data for different $B$ are plotted as a function of $T/D$, while in the main panel we demonstrate scaling collapse to a single universal curve when plotted as a function of the reduced parameter $T/B$. The scaling is remarkably robust, holding over a very wide range of parameters in both $T$ and $B$. This is the expected behavior for a free spin; as confirmed by direct comparison with the orange dotted line in Fig.~\ref{Fig:FB_Ising_Impurity_FS_Multiple}, corresponding to Eq.~\eqref{eq:freespin}.

Small deviations in the QSNR from the idealized free spin result, Eq.~\eqref{eq:freespin}, seen in our exact results for an Ising impurity in an explicit bath can be intuitively understood from mean-field theory. As shown in Appendix~\ref{app:ising}, within this simple approximation the impurity magnetization is given by, 
\begin{equation}
	\label{Eq:MFT_Imp_Mag1}
	\expval{\hat{S}_I^z} = - \frac{1}{2}\tanh{\left(\frac{B + {J^z}\expval{\hat{S}_0^z}}{2T}\right)} \;, 
\end{equation}
which is identical to the free spin result, except with a renormalized impurity field $B_I \to B + J^z\expval{\hat{S}_0^z}$. The bath magnetization $\expval{\hat{S}_0^z}$ must be obtained self-consistently, but is found to be rather small in the Ising case for all scenarios considered. We have verified numerically from the exact solution of the model that even at large $J^z \sim D$, the peak sensitivity attainable is still $\mathcal{Q}_{Max}\simeq \tfrac{2}{3}$, and indeed the full QSNR profile is essentially identical to the free spin result, but with a small field renormalization.

Our results indicate that the sensitivity of a quantum impurity spin coupled to a structured environment through an Ising interaction is analogous to the thermometric precision attainable for a free spin immersed in a memoryless bath~\cite{correa2015individual, Campbell_Percision_Thermo_QSL}. Furthermore, we find that this holds quite generally for structured baths with different DoS (see Appendix~\ref{app:bath}), with only minimal quantitative differences from the universal free-spin result. Our exact analytic solution of the Ising impurity model shows only a weak dependence on the bath DoS. 

This insensitivity of the probe to the structure of the environment, and the universal dependence of the QSNR on a single parameter $B/T$, together with the high $\mathcal{Q}_{max}$ attainable at \emph{any} temperature upon tuning the control field $B$, makes the Ising impurity a good thermometer. The Ising probe affords robust, reliable, and tuneable thermometric capabilities, irrespective of the host system being probed. However, the drawback is that this assumes that all other model parameters, and in particular the applied control-field strength, are precisely known~\cite{CampbellNJP2017}. Furthermore, such a setup does not allow for the generation of probe-environment entanglement, which spoils this ideal thermometric behavior.

 \begin{figure*}[t!]
	\begin{center}
    \includegraphics[width=1\linewidth]{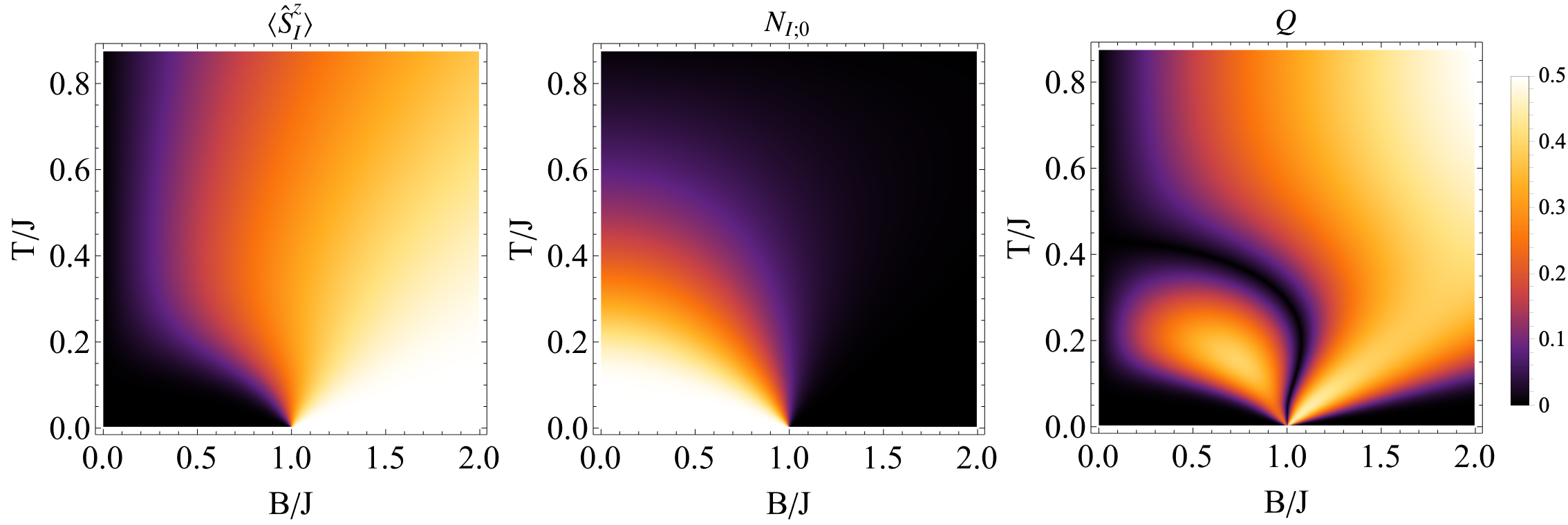}
    \end{center}
	\caption{Kondo impurity model in the narrow band limit $J \gg D$, given by Eq.~\eqref{Eq:Ham_NBL}. \textbf{(a)} Impurity magnetization $\expval{\hat{S}_I^z}$, \textbf{(b)} Entanglement negativity of the impurity $\mathcal{N}_{I;0}$, \textbf{(c)} Temperature sensitivity QSNR $\mathcal{Q}$. All plotted as a function of normalised field $B/J$ and temperature $T/J$.}
	\label{Fig:NBL}
\end{figure*}

\subsection{Kondo Coupling}\label{sec_Kondo_Coupling}
We now turn to the more realistic situation \cite{Hewson} in which the impurity probe and bath can become entangled. This arises due to the mutual spin-flip terms parameterized by $J^{\perp}$ in Eq.~\eqref{Eq:System_Hamiltonian}. In the following we take $J^z=J^{\perp}\equiv J$ and study the resulting isotropic Kondo model. We note that an impurity coupled to a fermionic bath by such a term has a natural thermalization mechanism since the impurity spin projection is no longer conserved. As noted above, the Kondo effect in such systems is a consequence of the low-temperature formation of a highly entangled many-body state. We explore the consequences of this for thermometry below. First, we motivate our discussion of the general Kondo case by consideration of two simple limits. 


\subsubsection{Large field limit}
The Kondo model in the limit of very large magnetic fields $B\gg J,D$ becomes simple since both the impurity and bath are polarized. The spin-flip terms in Eq.~\eqref{Eq:System_Hamiltonian} are suppressed and the impurity is effectively decoupled from the bath. The physics here is that of a free spin in a strong magnetic field.


\subsubsection{Narrow band limit}
Another simple and analytically-tractable limit of the Kondo model arises for strong coupling $J\!\gg\! D$. In this case the bare model is already close to the strong coupling renormalization group fixed point \cite{wilson1975renormalization}. Therefore the impurity forms a highly localized spin-singlet with the bath site to which it is coupled (one can view this as the limit where the Kondo entanglement cloud shrinks down in size to a single bath site). This is the narrow band limit, and we can approximate the bath by a single site. Taking $J^z=J^{\perp}\equiv J$ and $B_I=B_0\equiv B$, we write,
\begin{align}
    \label{Eq:Ham_NBL}
    \hat{H}_{NBL} = &\tfrac{1}{2} J \left [ \hat{S}^z_I(\hat{n}_{0\uparrow}-\hat{n}_{0\downarrow}) + \hat{S}^+_I c_{0\downarrow}^{\dagger} c_{0\uparrow}^{\phantom{\dagger}} + \hat{S}^-_I c_{0\uparrow}^{\dagger} c_{0\downarrow}^{\phantom{\dagger}}\right ] \nonumber \\
    &+ B \left [\hat{S}^z_I  + \tfrac{1}{2}(\hat{n}_{0\uparrow}-\hat{n}_{0\downarrow}) \right] \;.
\end{align}
The partition function of this two-site reduced model is simply $Z=4\cosh(B/2T)+e^{-J/4T}[1+2\cosh(B/T)] + e^{3J/4T}$. The impurity magnetization and negativity then follow as,
\begin{align}
    \expval{\hat{S}^Z_I} &= \frac{1}{2Z}\left [\left (e^{B/T}-1\right)\left(1+e^{B/T} + 2e^{(2B+J)/4T}\right)\right ] \\
    \mathcal{N}_{I;0} &= \frac{1}{2Z}\left [\sqrt{1+e^{2B/T}\left(e^{2B/T}+e^{2J/T}-1-2e^{J/T} \right)} -1-e^{2B/T}\right ]
\end{align}
From the impurity magnetization we obtain the QFI from Eq.~\eqref{Eq:QFI_Impurity_Model} and hence the QSNR from Eq.~\eqref{QSNR}. The results are shown in Fig.~\ref{Fig:NBL}. 

The behavior of this model is already richer than the free spin or Ising cases. Here we have a competition between the polarizing tendencies of the field $B$ and the propensity for spin-singlet formation favored by $J$. In the simplified two-site model, this results in a singlet-triplet level-crossing in the ground state from $\tfrac{1}{\sqrt{2}}(\ket{\!\uparrow\downarrow} - \ket{\!\downarrow\uparrow})$ for $B<J$ to $\ket{\downarrow\downarrow}$ for $B>J$. In the low-temperature limit, this quantum phase transition manifests as a discontinuous step in the impurity magnetization from $0$ to $-\tfrac{1}{2}$ at $B=J$, see Fig.~\ref{Fig:NBL}(a). This transition corresponds to a collapse of the entanglement between the impurity and the local bath site at $T=0$, Fig.~\ref{Fig:NBL}(b). However, at finite temperatures this transition is smoothed to a crossover: with increasing field strength $B$ the magnetization continuously increases, whereas the negativity decreases.

 \begin{figure*}[t!]
	\begin{center}\includegraphics[width=1.\linewidth]{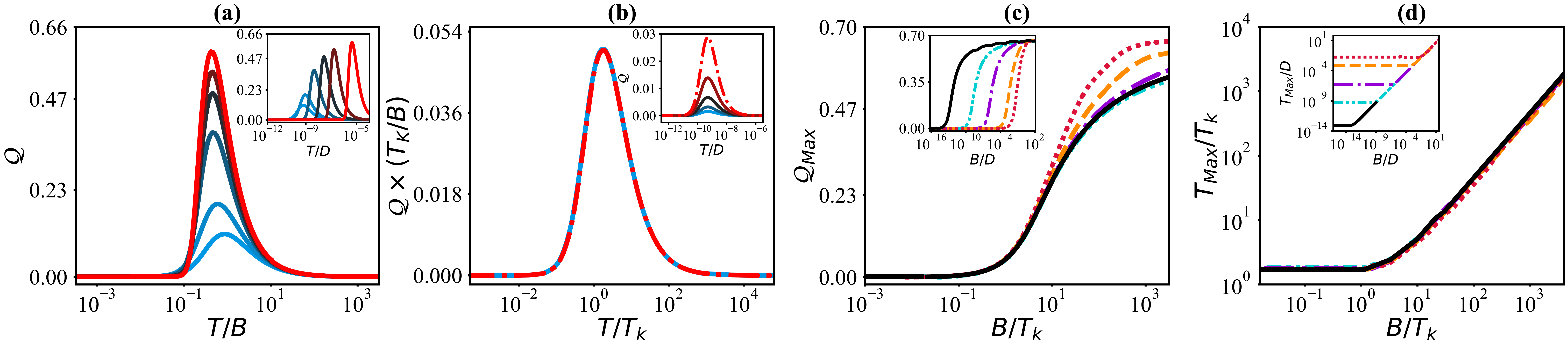}
	\end{center}
	\caption{Thermometry in the Kondo impurity model, with isotropic exchange coupling $J^z=J^{\perp}\equiv J$ for a flat conduction electron density of states. Calculations performed using NRG. \textbf{(a)} Quantum signal to noise ratio $\mathcal{Q}$ as a function of temperature $T$ for different field strengths $B$, plotted as $T/B$ in the main panel and $T/D$ in the inset. Shown for $B/D = 10^{-12},10^{-11},10^{-10},10^{-9},10^{-8},10^{-7}$ increasing from bottom-most-blue to topmost-red lines, for fixed $J=0.1D$. \textbf{(b)}  Universal QSNR curve in the Kondo regime $B\ll T_K$, plotted in terms of rescaled quantities $\mathcal{Q}\times (T_K/B)$ vs $T/T_K$. Shown for $B/D=10^{-12},10^{-11.5},10^{-11},10^{-10.5},10^{-10}$ for fixed $J=0.1D$ where we clearly see data collapse. Inset shows the same data plotted as the bare $\mathcal{Q}$ vs $T/D$.  \textbf{(c)} Peak sensitivity $\mathcal{Q}_{Max}$ as a function of $B/T_K$ (main panel) and $B/D$ (inset), for couplings $J/D = 0.07, 0.1, 0.15, 0.3, 0.5$ with corresponding Kondo temperatures $T_K/D \simeq 4 \times 10^{-14}, 3 \times 10^{-10}, 3 \times 10^{-7}, 4 \times 10^{-4}, 9 \times 10^{-3}$ denoted by the solid black, dot-dot dashed cyan, dot-dashed purple, dashed orange, and dotted red curves, respectively. \textbf{(d)} Corresponding temperature at which the sensitivity peaks, $T_{max}$, as a function of $B/T_K$ (main panel) and $B/D$ (inset).}
	\label{Fig:FB_Kondo}
\end{figure*}

This all gives rise to complex behavior in the thermometric sensitivity Fig.~\ref{Fig:NBL}(c). In particular, we note that the magnetization can be non-monotonic in $T$ due to the involvement of multiple competing states. From Eq.~\eqref{Eq:QFI_Impurity_Model} we conclude that the the QFI, and hence the QSNR, must develop a nontrivial nodal line in the $(B,T)$ plane, separating regions of enhanced thermometric sensitivity.

\subsubsection{General Kondo case}
Finally we consider the general case of the Kondo model at finite $B$ and $J$, with a spin-$\tfrac{1}{2}$ quantum impurity coupled to a full fermionic bath, where we restrict the metallic flat band case in this section and refer to Sec.~\ref{DoSsection} for a discussion of different DoS. The thermometric sensitivity is now affected by strong electron correlation effects. Here we have a new emergent scale, the Kondo temperature $T_K$, and the possibility of large-scale impurity-bath entanglement due to the formation of the macroscopic `Kondo cloud'. We also note that, unlike in the simplified two-site model, there is no quantum phase transition in the Kondo model as a function of field strength $B$ -- there is a smooth, universal crossover from the Kondo singlet ground state for $B\ll T_K$ to a polarized impurity for $B\gg T_K$. We use NRG \cite{wilson1975renormalization,bulla2008numerical,weichselbaum2007sum} to solve the Kondo model and obtain the impurity magnetization $\expval{\hat{S}_I^z}$, from which we extract the QSNR $\mathcal{Q}$ as a function of $T$ and $B$ for a given Kondo coupling $J$. The numerical results are plotted in Fig.~\ref{Fig:FB_Kondo} and discussed below.

We see that the strong electron correlations play an important role for the QSNR in the Kondo model, giving a nontrivial response beyond that of the simple free spin/Ising limit. In Fig.~\ref{Fig:FB_Kondo}(a) the applied field strength is seen to significantly affect the temperature probe sensitivity.
Only for very large fields $B$ (topmost, red line) do we see effective free spin/Ising behavior, with a characteristic peak in $\mathcal{Q}$ at $T\sim B$ approaching its maximum value of $\simeq \tfrac{2}{3}$. As the field strength decreases (see e.g.~lowermost, blue line), the maximum sensitivity attenuates and the details of the QSNR profile changes. Fundamentally, this is because the impurity spin-resolved populations in this regime are more robust to changes in the field, as the impurity is locked into a Kondo spin-singlet with bath electrons. The Kondo temperature $T_K$ sets the scale separating qualitatively different types of behavior: for $B\gg T_K$ Ising-like physics similar to Fig.~\ref{Fig:FB_Ising_Impurity_FS_Multiple} dominates, whereas for $B\ll T_K$ Kondo physics dominates. In this case, the peak position no longer scales with $B$ but instead becomes pinned around $T_K$, and the peak height scales with $B/T_K$.

Importantly, in the Kondo regime $B\ll T_K$, we find a completely universal behavior of the full QSNR profile, demonstrated in Fig.~\ref{Fig:FB_Kondo}(b). When properly rescaled as $\mathcal{Q}\times(T_K/B)$, we see scaling collapse to the same universal curve of all systems with different fields $B$ when plotted vs $T/T_K$. Despite qualitative similarities, we note that the shape of this universal curve is not the same as the free-spin or Ising result. This universality is significant, because we get the same result independently of underlying model parameters. In the following sections, we also show a stronger universality in the sense that the same response is obtained independently of the bath structure used (provided we still have a metallic system).  Thus, the Kondo probe can be used `model free', with very little information about the environment or the probe-environment coupling needed to make predictions about the temperature. 

However, the cost of achieving this universality is that the overall sensitivity in the Kondo-dominant regime $B\ll T_K$ is low. This is because the impurity and environment become highly entangled, and due to this high degree of non-classical correlation, the temperature sensing capabilities of the impurity are diminished. Intuitively, this follows from the fact that measurements confined to the impurity probe provide access to only a small amount of information about the entangled many-body Kondo state. To understand the role that probe-environment entanglement plays in determining the robustness and sensitivity of such an impurity thermometer, we study entanglement properties of the Kondo regime explicitly in Sec~\ref{sec_P_E_ent}.

The inset to Fig.~\ref{Fig:FB_Kondo}(b) shows that the QSNR peak \textit{position} (denoted $T_{Max}$) remains fixed at $T\sim T_K$ in the Kondo regime, while the peak height ($\mathcal{Q}_{Max}$) scales as $B/T_K$. By contrast, in the large-field Ising regime $B\gg T_K$ shown in Fig.~\ref{Fig:FB_Kondo}(a) we see $T_{Max}\sim B$ and $\mathcal{Q}_{Max}$ approaches the free-spin/Ising saturation value of $\sim \tfrac{2}{3}$. This non-linear thermometric performance of the probe is demonstrated in Fig.~\ref{Fig:FB_Kondo}(c), where the different lines correspond to different coupling strength $J$ values. The black curve for $J=0.07$ can be considered the universal curve, characterized by single-parameter scaling in $B/T_K$. In Fig.~\ref{Fig:FB_Kondo}(c) we see that $\mathcal{Q}_{Max}$ curves at larger $J$ progressively fold onto this universal curve at lower $B/T_K$. However non-universal deviations in $\mathcal{Q}_{Max}$ at large $B \sim J, D$ are expected. These are observed in practice at large $B/T_K$ when $T_K$ is itself large, since single-parameter scaling breaks down in this regime. We conclude that universal features of the probe appear only at small $T$ and $B$. 

In Fig.~\ref{Fig:FB_Kondo}(d) we show the behavior of $T_{Max}$, which similarly exhibits universal scaling in terms of $T_K$, and shows distinctive behavior in the Kondo and Ising regimes. We clearly see that once the Kondo effect is suppressed, i.e. $B\gg T_K$,  linear scaling of the ideal probe temperature is attained -- the same behavior exhibited by the free spin and Ising models. Finally, we note that high probe sensitivity, approaching that of the ideal free spin limit, can still be achieved at arbitrarily low temperatures in the Kondo system, simply by reducing the coupling $J$. Since small $J$ means a very small Kondo scale $T_K$ (recall that the dependence is exponential), one can go to the regime $B\gg T_K$ and see a peak in $\mathcal{Q}$ at temperatures $T\gg T_K$ that might still be very low in absolute terms. Here the field suppresses the Kondo spin-flip processes and hence the impurity-bath entanglement, yielding good thermometric capability. This can be at low absolute temperatures and still in the fully universal regime, provided $T_K \ll T,B \ll J,D$. 
Although the field strengths we have used for illustrative purposes in Fig.~\ref{Fig:FB_Kondo} are small in units of the conduction electron bandwidth $D$, we note that $D$ is typically a high-energy scale of order eV. The universal physics is set by the Kondo temperature $T_K$, and the relevant quantity in the Kondo regime is therefore $B/T_K$. To demonstrate scaling collapse of data for different $B$ and $J$ in Fig.~\ref{Fig:FB_Kondo} we have scanned $B$ over an exponentially wide range. However, we emphasize that in practice the universal regime can be realized at moderate $B$ provided $T_K$ is not very small. This simply corresponds to the regime of larger impurity-environment couplings $J$.


\begin{figure*}[t]
	\centering
	\includegraphics[width=1.\linewidth]{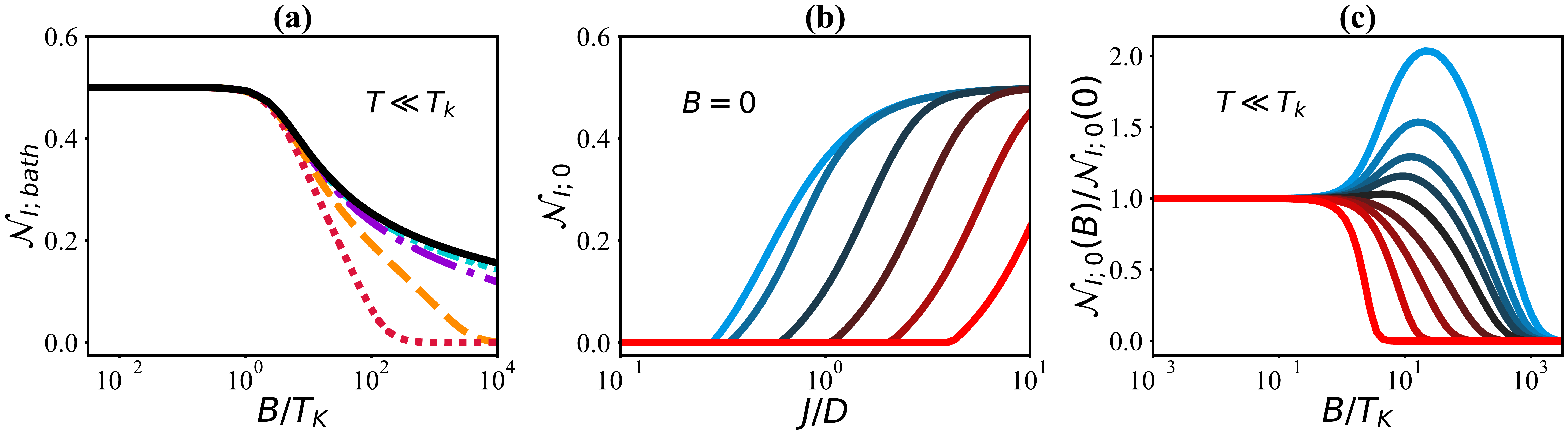}
	\caption{Entanglement negativity for the Kondo impurity model, with isotropic exchange coupling $J^z = J^\perp \equiv J$ for a flat conduction electron density of states. \textbf{(a)} Negativity of the impurity with the full bath $\mathcal{N}_{I;bath}$ in the limit $T \ll T_K$ as a function of $B/T_K$, for couplings $J/D = 0.07, 0.1, 0.15, 0.3, 0.5$ denoted by the solid black, dot-dot dashed cyan, dot-dashed purple, dashed orange, and dotted red curves, respectively. \textbf{(b)} Negativity of the impurity with the local bath site $\mathcal{N}_{I;0}$ in the zero-field limit $B = 0$, as a function of coupling strength $J$ for different temperatures $T/D = 0, 0.1, 0.5, 1, 2, 4$ increasing from leftmost blue to rightmost red lines. \textbf{(c)} Field-dependence of the negativity $\mathcal{N}_{I;0}$, normalized by its zero-field value, plotted for different couplings $J/D=0.3, 0.31, 0.32, 0.33, 0.35, 0.4, 0.5, 0.7, 1$ from rightmost blue to leftmost red, in the low-temperature limit $T\ll T_K$. 
 }
 \label{NegativityPlot}
\end{figure*}

\subsection{Probe-Environment entanglement}\label{sec_P_E_ent}
Next we consider in more detail the impurity-bath entanglement, focusing on the flat band Kondo model for metallic systems. This builds upon previous studies of the entanglement structure of Kondo systems~\cite{bayat2010negativity,shim2018numerical,lee2015macroscopic,han2022measure}, here generalized to the finite magnetic field case. We consider both the negativity between the impurity and the full bath $\mathcal{N}_{I;bath}$, as well as the negativity between the impurity and the local bath site to which it is connected (but still in the presence of the rest of the bath), denoted $\mathcal{N}_{I;0}$.

We consider first the low temperature limit $T \ll T_K$, which allows us to employ Eq.~\eqref{Eq:Neg_Full_Bath}, and compute the entanglement shared between the full bath and the impurity $\mathcal{N}_{I;bath}$, as a function of the applied control field $B$. This is plotted in  Fig.~\ref{NegativityPlot}(a) as a function of the rescaled parameter $B/T_K$ for different coupling strengths $J$. The universal curve, exhibiting single-parameter scaling in $B/T_K$ is obtained for small $T_K$, meaning in practice small coupling $J$, see solid black line in Fig.~\ref{NegativityPlot}(a). At larger $J$ non-universal deviations are observed at larger $B$ due to band-edge effects. However, the trend in all cases is the same: an applied field on the order of the Kondo temperature suppresses impurity-bath entanglement. For fields $B \ll T_K$ however, the impurity-bath negativity is maximised, meaning that the impurity in maximally entangled with the full bath, independent of the coupling strength $J$. This is attributed to the spin flip terms being the predominant process over the small degree of spin polarisation caused by finite $B$, and the resulting Kondo effect. In this regime the reduced temperature sensitivity of the probe shown in Fig.~\ref{Fig:FB_Kondo}(c) can be attributed to this strong entanglement. Conversely, in the opposite limit, $B \gg T_K$, the negativity starts to decay as the applied field becomes the dominant term. Here the trade-off between the competing effects enhances the thermometric capacity of the probe, while still producing universal behavior. Further increasing the field strength, the entanglement decays and the impurity's sensing capabilities approach that of the Ising/free-spin limit. The qualitative behavior at finite temperatures is expected to be similar.

In the universal regime, the impurity-bath entanglement generated by the Kondo effect is highly non-local~\cite{v2020observation, bayat2010negativity,lee2015macroscopic,shim2018numerical,kim2021universal}, extending in real-space throughout the Kondo cloud, which involves a macroscopic number of bath sites~\cite{mitchell2011real}. This entanglement cloud shrinks in size as the coupling strength $J$ (and hence the Kondo temperature $T_K$) increases. This is demonstrated indirectly in Fig.~\ref{NegativityPlot}(b), where we show the negativity between the impurity and just the local bath site to which it is connected $\mathcal{N}_{I;0}$, as a function of coupling strength $J$, at zero field, $B=0$. At zero temperature (in practice $T\ll T_K$), the impurity is seen to become maximally entangled with the local bath site when $J\gtrsim 1$, see leftmost blue line. Due to the monogamy of entanglement, this implies that the impurity is disentangled with the rest of the bath; this is the narrow band limit considered previously, where the Kondo cloud essentially shrinks down to a single site. At smaller $J$, the entanglement shared between the impurity and the local bath site is much smaller. But since for $T\ll T_K$ and $B=0$ the entanglement between the impurity and the full bath is large, Fig.~\ref{NegativityPlot}(a), we can infer that the bulk of the entanglement is with sites further away from the impurity. By increasing the temperature, from leftmost blue to rightmost red lines in Fig.~\ref{NegativityPlot}(b), we see a similar trend, but with a delayed onset of the maximum local entanglement with $J$. 

In Fig.~\ref{NegativityPlot}(c) we examine the subtle interplay between finite $B$ and $J$ in the local entanglement. We plot $\mathcal{N}_{I;0}$ at finite $B$ in the low-temperature regime, normalized by its zero-field value, with the different lines corresponding to different $J$, increasing from rightmost blue to leftmost red. Here we see non-monotonic behavior in the local entanglement at smaller $J$, see e.g.~blue line for $J=0.3$. In this case, the Kondo entanglement cloud at $T=B=0$ is large, and only a small amount of the full impurity-bath entanglement is shared with the local bath site. But by turning on the field to $B\sim T_K$, the Kondo effect starts to be suppressed. The many-body Kondo cloud begins to evaporate and although the overall impurity-bath entanglement decreases, see Fig.~\ref{NegativityPlot}(a), it is also redistributed in real-space in a nontrivial way. The local spin-exchange process begins to dominate over the many-body Kondo effect and the entanglement shared between the impurity and the local bath site grows with increasing $B$. For larger fields $B\gg T_K$ however, the impurity becomes polarized and the entanglement collapses when $B\sim J$. We note that when the bare coupling $J$ is large, e.g.~red line for $J=1$, the Kondo cloud is already small and a high proportion of the full impurity-bath entanglement is local. In this case, we approach a step-function decay of the entanglement on increasing $B$. This is precisely the behavior expected in the narrow-band limit shown in \ref{Fig:NBL}(b), where at low $T$ we found that $\mathcal{N}_{I;0}$ goes from its maximum $\tfrac{1}{2}$ for $B<J$ to $0$ for $B>J$. In the true Kondo model, this transition is smoothed to a crossover, but one that becomes progressively sharper with increasing $J$. The unusual non-monotonic behavior appears only in the universal regime at smaller $J$ due to many-body effects.

\section{Dependence on Environmental Spectral Features}
\label{DoSsection}
In the previous sections we focused on an impurity probe embedded in a metallic environment with a featureless density of states. Here we generalize to structured environments, characterized by their spectral properties via the local DoS $\rho(\omega)$ at the probe location. Different physical systems will of course have different DoS features, which depend on the structure, geometry, dimensionality, and ultimately the chemistry of the host material, through its band structure. Although $\rho(\omega)$ for real systems will be complicated, for the purposes of the metrological quantum impurity problem we can broadly classify them into three families according to their low-energy behavior: {\it (i)} metals with constant DoS $\rho(\omega)\sim \rho_0$ at low energies; {\it (ii)} semi-metals with a low-energy pseudogap vanishing DoS $\rho(\omega)\sim |\omega|^r$ with $r>0$; {\it (iii)}  materials with a low-energy power-law diverging DoS $\rho(\omega)\sim |\omega|^r$ with $-1<r<0$. The impurity response in each case is known to be quite different \cite{Hewson,gonzalez1998renormalization,mitchell2013kondodiv,*mitchell2013quantum}; nevertheless distinct systems within a family share the same universal features. Representative material examples of the three classes are gold \cite{costi2009kondo} (metallic), graphene \cite{neto2009electronic,fritz2013physics} (semi-metal pseudogap with $r=1$), and magic-angle twisted bilayer graphene (TBG) \cite{tarnopolsky2019origin,lisi2021observation,yuan2019magic} (power-law diverging DoS with $r=-1/4$). Note that we can study systems with a finite electronic bandwidth $D$, or with an infinite bandwidth provided $\rho(\omega)$ is still normalizable. Further details of the specific DoSs used for our calculations in this section can be found in Appendix~\ref{app:bath}.

As alluded to previously, for the Ising impurity model ($J^{\perp}=0$), we find that there is essentially no dependence of the QSNR on the bath DoS $\rho(\omega)$, and the free spin result Eq.~\eqref{eq:freespin} pertains except for very large $J \gtrsim D$, where small deviations are observed. The structure of the bath in this case does not play a significant role in part due to the fact that the impurity and bath do not become entangled.

\begin{figure}[t]
	\begin{center}\includegraphics[width=0.8\linewidth]{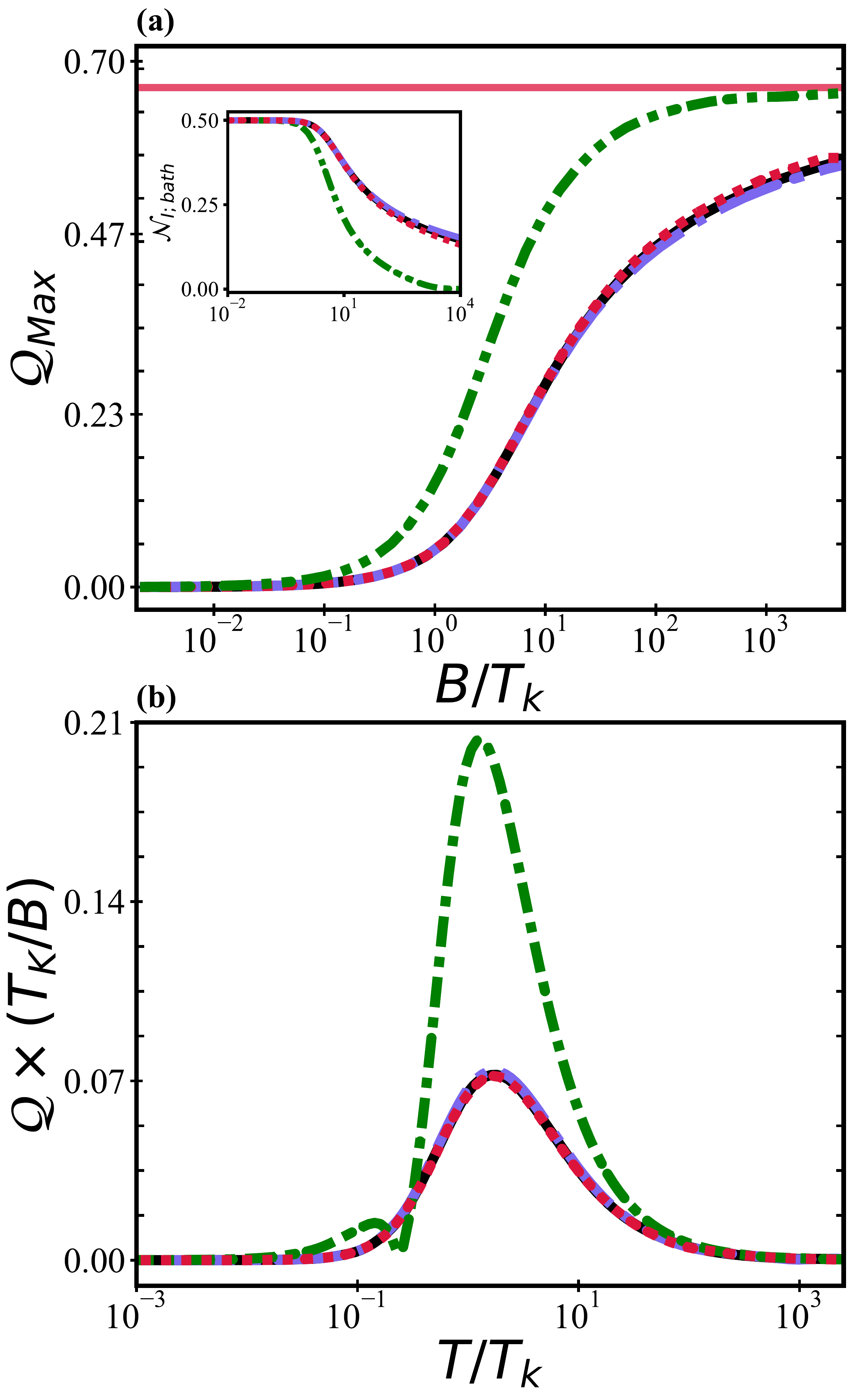}
	\end{center}
	\caption{Universal behaviour of Kondo impurity probes in structured baths with coupling $J = 0.1D$ in terms of the conduction bandwidth $D$. \textbf{(a)} Peak sensitivity $\mathcal{Q}_{Max}$ as a function of $B/T_K$ for bath systems with DoS corresponding to flat band (black line), nanowire (red dot), Gaussian (blue dashed), and diverging DoS (green dashed) with corresponding Kondo temperatures $T_K/D \simeq 3 \times 10^{-10}, 2 \times 10^{-8}, 2 \times 10^{-12}, 5 \times 10^{-4}$. Red line shows the ideal free spin/Ising result. The inset shows the corresponding impurity-bath entanglement negativity decay $\mathcal{N}_{I;bath}$ in the $T \ll T_K$ limit. \textbf{(b)} Rescaled QSNR $\mathcal{Q}\times(T_K/B)$ vs $T/T_K$ for the same systems.}
	\label{Fig:Different_DOS}
\end{figure}
The situation is quite different in the Kondo case (finite $J^{\perp}$) because the bath DoS strongly influences the development of impurity-bath entanglement via the Kondo effect. 
Relative to a standard metallic system, the Kondo temperature $T_K$ for a system with diverging DoS (such as TBG) is strongly enhanced~\cite{mitchell2013kondodiv,*mitchell2013kondo}. This means that the impurity probe and the bath become strongly entangled on much higher temperature scales than in metallic systems -- and therefore that a much higher field $B$ must be applied to suppress the Kondo effect and recover Ising-like sensitivity.  
However, even when rescaled in terms of $B/T_K$, we expect differences between metallic DoS and diverging DoS systems, since they belong to different quantum impurity universality classes.  To demonstrate this, in Fig.~\ref{Fig:Different_DOS}(a) we show the peak sensitivity $\mathcal{Q}_{Max}$ vs $B/T_K$ in the universal regime $B,T \ll J,D$, for different systems with $\rho(\omega)$ corresponding to the metallic flat band (black line), metallic nanowire (red dotted), metallic Gaussian (blue dot-dash), and TBG power-law diverging DoS (green dashed). The three metallic systems have the same universal behaviour, independent of the detailed structure of the DoS, when rescaled in terms of $B/T_K$. The power-law diverging DoS also gives a universal response, but one that is different from the metallic case, saturating to the Ising limit $\mathcal{Q}_{Max}\simeq \tfrac{2}{3}$ (red line) more rapidly than the metals as $B/T_K$ is increased. 
The differences between metallic and diverging DoS are also evident in Fig.~\ref{Fig:Different_DOS}(b) where we show the full QSNR profile for the same systems. We plot rescaled quantities to give the universal curves, $\mathcal{Q}\times (T_K/B)$ vs $T/T_K$. As expected we see distinct behavior for the two different families (metallic and diverging DoS), but the same behaviour for different systems within the same family (different $\rho(\omega)$ and $J$). 

We can develop a more fundamental understanding of the qualitative differences that arise in the impurity probes' thermometric response when in contact with different environments by considering the impurity-bath entanglement via the negativity, $\mathcal{N}_{I;bath}$, shown in the inset of Fig.~\ref{Fig:Different_DOS}(a). Here we see that the TBG and metallic DoS have distinct decay rates on applying a field $B$ (even when plotted as $B/T_K$), with the former dropping off more rapidly than the latter -- a consequence of the spin becoming more easily polarised for the TBG case.

The Kondo model is a realistic model of quantum impurity probes in real fermionic host systems, and provides a natural thermalization mechanism through the $J^{\perp}$ spin-flip scattering processes. However, these same spin-flip scattering processes also lead to the Kondo effect and hence macroscopic impurity-bath entanglement in metallic systems, which strongly suppresses thermometric sensitivity. We therefore conclude the that Kondo impurity probe is extremely robust for $B\ll T_K$, giving the same universal `model-free' behvior, independent of the bare parameters and bath DoS -- provided the systems considered are within the same impurity universality class. Despite the reduction in absolute terms due to the strong entanglement established in this regime, this nevertheless demonstrates a remarkable advantage over other thermometric schemes since knowledge of the precise details of the host system being probed are not required.

Finally, we remark on the interesting case for the power-law vanishing case of a graphene bath (results not shown). The depleted DoS around the Fermi energy suppresses the Kondo effect~\cite{fritz2013physics} and therefore the impurity remains a free local moment down to $T\!=\!0$. The thermometric sensitivity of a spin qubit probe in graphene is therefore the same as that of a free spin in an implicit thermal environment (Gibbs state), Eq.~\eqref{eq:freespin}. Thus, an impurity in graphene can still thermalize, but the low-energy linear pseudogap in the DoS prevents the Kondo effect from forming and hence the arrests the development of strong impurity-bath entanglement. In turn, this means that in graphene, we obtain near-perfect (free spin/Ising-like) thermometry. Our findings therefore suggest that a magnetic impurity embedded on a large graphene flake could together act as an excellent quantum thermometer, assuming that the graphene flake could itself thermalize with its surroundings.


\section{Conclusions}
We have examined the thermometric performance of quantum spin-$\tfrac{1}{2}$ impurity probes immersed in structured fermionic quantum environments. Assuming that the impurity and environment are in thermal equilibrium, we have carefully studied the role that the nature of the impurity-environment coupling, the resulting entanglement, and the environment's spectral properties play in the achievable thermometric precision. For an Ising-type coupling between the probe and environment, no entanglement is generated. In this case, we find that impurities act as versatile and sensitive probes, achieving peak sensitivities comparable to those achievable for an idealized two-level system. Furthermore, this sensitivity is independent of the specific structure of the environment being probed, and the temperature at which peak sensitivity is achieved scales linearly with the applied control field. However, such a setup has no intrinsic thermalization mechanism, and is arguably an oversimplified model for realistic impurity probes \cite{Hewson}. 

On the other hand, allowing for spin-flip exchange coupling terms between the impurity probe and the quantum environment, leads to the build up of strong probe-environment entanglement at lower temperatures, and we show this has a dramatic effect on the  thermometric response. In this Kondo impurity model, an emergent low-energy scale $T_K$, characterizing the onset of strong correlations, distinguishes two separate regimes. When the control field is large, $B\gg T_K$, probe-environment entanglement is suppressed and the Ising interaction dominates. Then the thermometic capability of the probe is again comparable to that of the idealized two-level system. For $B\ll T_K$, strong probe-environment entanglement generated by the Kondo effect leads to reduced overall sensitivity. However, in this regime we uncover a \textit{universal} probe response in the thermal QSNR, independent of bare model parameters and microscopic details. This indicates that by sacrificing sensitivity we can exploit many-body effects to achieve sensors that do not require knowledge of system parameters and operate `model free'. Indeed, this is a particularly relevant point in thermometric protocols where often one implicitly assumes that the temperature is the only parameter to be estimated, and all other terms entering the Hamiltonian are known precisely~\cite{CampbellNJP2017}. Our results demonstrate that this requirement can be circumvented by leveraging the universal features of many-body systems. 

Finally, as an outlook, we comment that the Kondo model studied here is also the low-energy effective model describing experimental semiconductor quantum dot systems \cite{goldhaber1998kondo,iftikhar2018tunable,bashir2021monolithic,*blokhina2019cmos,piquard2023observation}, meaning that our results for temperature sensing may be relevant to existing quantum nanoelectronics devices. The Kondo regime can be accessed experimentally in quantum dot devices, with data for different systems collapsing to a single universal curve when plotted in terms of $T/T_K$. This is possible using experimental temperatures and control fields because the Kondo temperature $T_K$ is highly sensitive to the dot-lead coupling strength, and in practice can be tuned \emph{in-situ} from $\sim 10^{-6}$K to $10^{+2}$K by varying gate voltages, see e.g.~\cite{iftikhar2018tunable}. Furthermore, the key observable discussed in this work -- the probe magnetization -- can be measured experimentally in the Kondo regime in such quantum dot systems, see e.g.~Ref.~\cite{piquard2023observation}.

 A further interesting direction is to extend the study to more complex impurity systems, including multi-impurity setups \cite{pouse2023quantum}, those exhibiting quantum criticality~\cite{AbolPRB}, and/or to systems out of equilibrium~\cite{MitchisonPRL}.


\acknowledgments
We acknowledge fruitful exchanges with Sindre Brattegard, Gabriele De Chiara, Gabriel Landi, Mark Mitchison, and Seung-Sup B. Lee. This work was supported by Equal1 Labs (GM), the Science Foundation Ireland Starting Investigator Research Grant “SpeedDemon” No. 18/SIRG/5508 and the John Templeton Foundation Grant ID 62422 (SC), and the Irish Research Council through the Laureate Award 2017/2018 grant IRCLA/2017/169 (AKM).


\appendix

\section{Structured fermionic environments}\label{app:bath}
In this work we consider impurity probes for explicit fermionic environments. We assume that the environment consists of a bath of non-interacting electrons in the thermodynamic limit, described by the diagonal Hamiltonian $\hat{H}_{\textit{bath}}=\sum_{k,\sigma} \epsilon_k \hat{c}_{k\sigma}^{\dagger} \hat{c}_{k\sigma}^{\phantom{\dagger}}$.
However, rather than specifying a particular tight-binding model, dispersion, or band-structure for the bath, we work directly with its local DoS $\rho(\omega)$, at the impurity location.  $\rho(\omega)$ is in general continuous and structured. It is related to the retarded electron Green's function of the free bath at the impurity probe location, via
$\rho(\omega)=-\tfrac{1}{\pi}{\rm Im}~G_{00}^0(\omega)$. Here, the real-frequency Green's function 
$G_{00}^0(\omega)=\langle\langle c_{0\sigma}^{\phantom{\dagger}} ; c^{\dagger}_{0\sigma}\rangle\rangle^0$ is the Fourier transform of the real-time propagator $G_{00}^0(t) = -i\theta(t)\langle \{ \hat{c}_{0\sigma}^{\phantom{\dagger}}(t) ,\hat{c}_{0\sigma}^{\dagger}(0)\} \rangle^0$, and 
$\hat{c}_{0\sigma}^{(\dagger)}=\sum_k \xi_k \hat{c}_{k\sigma}^{(\dagger)}$ where $\xi_k$ is the weight of a single-particle bath eigenstate $k$ at the impurity position.

We consider the following paradigmatic cases for the DoS:
\begin{align}
    \rho_{\textit{flb}}(\omega) &= \rho_0 ~\theta(1-|\omega/D|) ~~ &:&~ \text{Flat band (metallic)} \label{eq:dos_flb}\\
    \rho_{\textit{nw}}(\omega) &= \rho_0 \sqrt{1-(\omega/D)^2} ~~ &:&~ \text{Nanowire (metallic)}\label{eq:dos_nw}\\
    \rho_{\textit{gau}}(\omega) &= \rho_0\exp[-(\omega/D)^2] ~~ &:&~ \text{Gaussian (metallic)}\label{eq:dos_gau}\\
    \rho_{\textit{grph}}(\omega) &\sim \rho_0 ~|\omega/D| ~~ &:&~ \text{Graphene (semi-metal)}\label{eq:dos_grph}\\
    \rho_{\textit{TBG}}(\omega) &\sim \rho_0 ~|\omega/D|^{-1/4} ~~ &:&~ \text{TBG (diverging)}\label{eq:dos_tblg}
\end{align}
In each case, $\rho_0$ is chosen such that the corresponding DoS correctly normalizes to 1. Eqs.~\ref{eq:dos_flb}-\ref{eq:dos_gau} describe metallic systems, with the flat band and nanowire having finite hard bandwidths $D$, while the Gaussian case has an infinite bandwidth but still a characteristic width $D$. We obtain the semi-elliptical nanowire DoS from a semi-infinite tight-binding chain model. In Eq.~\ref{eq:dos_grph} graphene is our chosen representative of the pseudogap DoS class, with its low-energy Dirac cone spectrum (in our calculations for graphene we use the full DoS of the hexagonal tight-binding lattice \cite{neto2009electronic,fritz2013physics}). Finally, we consider magic-angle twisted bilayer graphene in Eq.~\ref{eq:dos_tblg}, which has a power-law diverging DoS at low-energies due to flat bands in the full band-structure coming from the higher-order van Hove singularity in that material \cite{tarnopolsky2019origin,lisi2021observation,yuan2019magic,shankar2023kondo}.

In terms of a magnetic impurity embedded in these host materials,  metallic systems have a Kondo temperature $T_K \sim D e^{-1/\rho_0 J}$ \cite{Hewson}, while the Kondo effect in TBG is strongly boosted, $T_K\sim D (\rho_0 J)^4$ \cite{shankar2023kondo}. By contrast there is no Kondo effect in neutral graphene, $T_K=0$, due to the depleted DoS.


\section{Solution of the Kondo impurity model}\label{app:kondo}
When the spin-flip terms embodied by finite $J^{\perp}$ in Eq.~\ref{Eq:System_Hamiltonian} are included, the model becomes a highly nontrivial quantum many-body problem for which no generic analytic solution exists. As such, sophisticated numerical methods such as Wilson's NRG technique are needed to obtain the full solution. Note that basic mean-field methods totally fail to capture the physics of the Kondo model, unlike the Ising case.


\subsection{Numerical Renormalization Group}
NRG is a numerically-exact, non-perturbative technique for solving quantum impurity type problems \cite{wilson1975renormalization,bulla2008numerical}. It was originally designed to deal with the metallic Kondo problem, but has since been generalized to impurities in arbitrary structured baths  \cite{chen1995kondo,*bulla1997anderson,bulla2008numerical}. Here we use full-density-matrix NRG \cite{weichselbaum2007sum} to obtain the (local) impurity magnetization of the Kondo model in the presence of impurity and bath magnetic fields, at finite temperatures, as well as to calculate the impurity entanglement negativity \cite{shim2018numerical}.

The NRG method involves the following steps \cite{wilson1975renormalization,bulla2008numerical}.\\ 
(i) The bath DoS $\rho(\omega)$ is divided up into intervals on a logarithmic grid according to the discretization points $\pm D \Lambda^{-n}$, where $D$ is the bare conduction electron bandwidth, $\Lambda>1$ is the NRG discretization parameter, and $n=0,1,2,3,...$. The continuous electronic density in each interval is replaced by a single pole at the average position with the same total weight, yielding $\rho^{\rm disc}(\omega)$.\\
(ii) The conduction electron part of the Hamiltonian $H_{\textit{bath}}$ is then mapped onto a semi-infinite tight-binding chain (called the `Wilson chain'),
\begin{eqnarray}
    H_{\textit{bath}} \to H_{\textit{bath}}^{\rm disc} = \sum_{\sigma}\sum_{n=0}^{\infty} &t_n^{\phantom{\dagger}} \left ( f_{n,\sigma}^{\dagger}f_{n+1,\sigma}^{\phantom{\dagger}}+  f_{n+1,\sigma}^{\dagger}f_{n,\sigma}^{\phantom{\dagger}}\right ) \;,
\end{eqnarray}
where the Wilson chain coefficients $\{t_n\}$ are determined such that the DoS at the end of the chain reproduces exactly the discretized host DoS, that is $-\tfrac{1}{\pi}{\rm Im}~\langle\langle f_{0,\sigma}^{\phantom{\dagger}} ; f_{0,\sigma}^{\dagger}\rangle\rangle = \rho^{\rm disc}(\omega)$. For simplicity we have assumed particle-hole symmetry in the free bath here. Due to the logarithmic discretization, the Wilson chain  parameters decay roughly exponentially down the chain, $t_n \sim \Lambda^{-n/2}$. However the detailed form of the $t_n$ encode the specific host DoS \cite{bulla2008numerical}.\\ 
(iii) The impurity is coupled to site $n=0$ of the Wilson chain. We define a sequence of Hamiltonians $H_N$ comprising the impurity and the first $N+1$ Wilson chain sites,
\begin{equation}\label{eq:Hn}
\begin{split}
    H_N=&J \hat{\vec{S}}_I \cdot \hat{\vec{S}}_0 + B\left( \hat{S}^z_I + \hat{S}^z_0\right )+ \sum_{\sigma}\sum_{n=0}^{N-1}  t_n^{\phantom{\dagger}} \left ( f_{n,\sigma}^{\dagger}f_{n+1,\sigma}^{\phantom{\dagger}}+  f_{n+1,\sigma}^{\dagger}f_{n,\sigma}^{\phantom{\dagger}}\right )  \;,
\end{split}
\end{equation}
where we assume $J^z=J^{\perp}\equiv J$ 
 and $B_I=B_0 \equiv B$ for simplicity. We now define the recursion relation,
\begin{equation}\label{eq:recursion}
    H_{N+1}=H_{N}+\sum_{\sigma}t_N^{\phantom{\dagger}} \left ( f_{N,\sigma}^{\dagger}f_{N+1,\sigma}^{\phantom{\dagger}}+f_{N+1,\sigma}^{\dagger}f_{N,\sigma}^{\phantom{\dagger}}\right ) \;,
\end{equation}
such that the full (discretized) model is obtained as $H^{\rm disc}=\lim_{N\to \infty} H_N$. \\
(iv) Starting from the impurity, the chain is built up successively by adding Wilson chain sites using the recursion, Eq.~\ref{eq:recursion}. At each step $N$, the intermediate Hamiltonian $H_N$ is diagonalized, and only the lowest $N_s$ states are retained to construct the Hamiltonian $H_{N+1}$ at the next step, with the higher energy states being discarded. With each iteration we therefore focus on progressively lower energy scales. Furthermore, the iterative diagonalization and truncation procedure can be viewed as an RG transformation \cite{wilson1975renormalization}, $H_{N+1}=\mathcal{R}[H_N]$.\\
(v) The partition function $Z_N$ can be calculated from the diagonalized Hamiltonian $H_N$ at each step $N$. Wilson showed \cite{wilson1975renormalization} that thermodynamic properties obtained from $Z_N$ at an effective temperature $T_N \sim \Lambda^{-N/2}$ accurately approximate those of the original undiscretized model at the same temperature. The sequence of $H_N$ can therefore be viewed as coarse-grained versions of the full model, which faithfully capture the physics at progressively lower and lower temperatures. Useful information is therefore extracted from each step, and physical observables can be obtained at essentially any temperature.

In this work, we use an NRG discretization parameter $\Lambda=2$ and retain $N_s=3000$ states at each step of the calculation. Total charge and spin projection abelian quantum numbers are exploited in the block diagonalization procedure.


\section{Solution of the Ising impurity model}\label{app:ising}
For an Ising impurity probing a fermionic environment, the full Hamiltonian is given by Eq.~\ref{Eq:System_Hamiltonian} with $J^{\perp}=0$. Even though this is an interacting quantum many-body problem (its fermionic representation involves quartic terms), it is trivially integrable and therefore exactly solvable because the impurity spin is conserved, $[\hat{H},\hat{S}_I^z]=0$. This means that exact eigenstates are separable (product) states, $|\psi\rangle=|S^z_I\rangle_I \otimes |\phi\rangle_B$, with corresponding energies $E={_B}\langle \phi| \hat{H}_{\textit{bath}}|\phi \rangle_B + B_I S^z_I + (B_0+J^z S^z_I){_B}\langle \phi| \hat{S}^z_0|\phi \rangle_B$. Given that $\hat{S}^z_0=\tfrac{1}{2}[\hat{n}_{0\uparrow} - \hat{n}_{0\downarrow}]$ and that for a given impurity spin projection $S^z_I$ the bath electrons with spin $\sigma=\uparrow$ and $\downarrow$ are strictly decoupled, the effect of the impurity on the conduction electrons can be entirely captured with a modified boundary potential added to the free $\hat{H}_{\textit{bath}}$. We can therefore write,
\begin{equation}\label{eq:Hising_sep}
    \hat{H} =  B_I^{\phantom{z}} \hat{S}^z_I + \sum_{S^z_I,\sigma} |S^z_I\rangle  \hat{H}_{\textit{bath}}^{S^z_I,\sigma} \langle S^z_I | \;,
\end{equation}
where, 
\begin{equation}\label{eq:Hbath_mod}
    \hat{H}_{\textit{bath}}^{S^z_I,\sigma}= \epsilon_{\textit{eff}}^{S^z_I,\sigma} \left( \hat{n}^{\phantom{\dagger}}_{0\sigma}\right ) + \sum_k \epsilon^{\phantom{\dagger}}_k \hat{c}_{k\sigma}^{\dagger} \hat{c}_{k\sigma}^{\phantom{\dagger}} 
    \;,
\end{equation}
and $\epsilon_{\textit{eff}}^{S^z_I,\sigma}=B_0^{\phantom{z}} S^z_0 + J^z S^z_I \sigma$ is the effective boundary potential ($\sigma=\pm \tfrac{1}{2}$ for $\uparrow/\downarrow$ electron spin). Here $\hat{n}_{0\sigma}=\hat{c}_{0\sigma}^{\dagger} \hat{c}_{0\sigma}^{\phantom{\dagger}}$ is the number operator for the local bath site to which the impurity is coupled, and $c_{0\sigma}=\sum_k \xi_k c_{k\sigma}$ as before. Each of the four effective models $\hat{H}_{\textit{bath}}^{S^z_I,\sigma}$ are simple free-fermion problems that are easily solved. Here we are interested in calculating the QFI for the Ising impurity system in an arbitrary structured fermionic environment, characterized by its free DoS, $\rho(\omega)$. We therefore develop a Green's function method to solve the model exactly, which allows us to work directly in the thermodynamic limit of the bath, with $\rho(\omega)$ as the input. The impurity magnetization $\langle S^z_I\rangle$ is calculated exactly, and the QFI follows immediately from Eq.~\ref{Eq:QFI_Impurity_Model}.

The impurity magnetization is obtained from the free energy, $\langle S^z_I\rangle=\partial F/\partial B_I$, with $F=-T\ln Z$ where $Z$ is the full partition function. Since eigenstates are labelled by $S^z_I$, we can decompose $Z=\sum_{S^z_I} Z^{S^z_I}$. From Eq.~\ref{eq:Hising_sep} it follows that $Z^{S^z_I}=Z^{S^z_I}_I\prod_{\sigma}Z^{S^z_I,\sigma}_0$ where $Z^{S^z_I}_I=e^{- B_I^{\phantom{z}}S^z_I/T}$ and $Z^{S^z_I,\sigma}_B$ is the partition function calculated from $\hat{H}_{\textit{bath}}^{S^z_I,\sigma}$ in Eq.~\ref{eq:Hbath_mod}. The impurity magnetization follows as,
\begin{equation}\label{eq:mag}
    \langle S^z_I\rangle= \frac{1}{2} \left ( \frac{Z^{\uparrow} - Z^{\downarrow}}{Z^{\uparrow} + Z^{\downarrow}}\right) \;.
\end{equation}

To make progress we must calculate $Z^{S^z_I,\sigma}_B$. Since $\hat{H}_{\textit{bath}}^{S^z_I,\sigma}$ is quadratic, we may use,
\begin{equation}\label{eq:zfunc}
    \ln Z^{S^z_I,\sigma}_B = \int d\omega~\rho_{\rm tot}^{S^z_I,\sigma}(\omega) \times \ln[1+e^{-\omega/T}] \;,
\end{equation}
where $\rho_{\rm tot}^{S^z_I,\sigma}(\omega)$ is the \emph{total} (orbital-summed) DoS of $\hat{H}_{\textit{bath}}^{S^z_I,\sigma}$. We separate this into a contribution $\rho_{\rm tot}^0(\omega)$ from the free bath (evaluated for $\epsilon_{\textit{eff}}^{S^z_I,\sigma}=0$) and a contribution $\Delta\rho_{\rm tot}^{S^z_I,\sigma}(\omega)$ due to the introduction of the boundary potential (evaluated for finite $\epsilon_{\textit{eff}}^{S^z_I,\sigma}$), such that $\rho_{\rm tot}^{S^z_I,\sigma}(\omega)=\rho_{\rm tot}^0(\omega)+\Delta\rho_{\rm tot}^{S^z_I,\sigma}(\omega)$. Since the contribution to $Z^{S^z_I,\sigma}_B$ coming from $\rho_{\rm tot}^0(\omega)$ does not depend on $S^z_I$ or $\sigma$, it cancels in the magnetization calculation, Eq.~\ref{eq:mag}, and therefore we do not need to evaluate it explicitly. We obtain $\Delta\rho_{\rm tot}^{S^z_I,\sigma}(\omega)=-\tfrac{1}{\pi}{\rm Im}~\Delta G_{\rm tot}^{S^z_I,\sigma}(\omega)$ from the change in the total (orbital-summed) bath Green's function due to the boundary potential.

The free bath DoS at the impurity position $\rho(\omega)=-\tfrac{1}{\pi}{\rm Im}~G_{00}^0(\omega)$ is related to the retarded local Green's function $G_{00}^0(\omega)=\langle\langle c_{0\sigma}^{\phantom{\dagger}} ; c^{\dagger}_{0\sigma}\rangle\rangle^0$, evaluated for the decoupled $H_{\textit{bath}}$, and can have arbitrary structure. We may write this Green's function as $G_{00}^0(\omega)=1/[\omega+i0^+-\Delta(\omega)]$ in terms of an auxiliary hybridization function $\Delta(\omega)$. Including the boundary potential, the Dyson equation yields,
\begin{equation}\label{eq:dyson}
    \left [G_{00}^{S^z_I,\sigma}(\omega)\right ]^{-1} = \left [G_{00}^0(\omega)\right ]^{-1} - \epsilon_{\textit{eff}}^{S^z_I,\sigma} \;.
\end{equation}
The difference in the local Green's functions at the impurity position due to the boundary potential is then $\Delta G_{00}^{S^z_I,\sigma}(\omega) = G_{00}^{S^z_I,\sigma}(\omega) - G_{00}^0(\omega)$. Finally, we use Green's function equations of motion methods \cite{zubarev1960double,Hewson} to express the
\textit{total} change in the bath Green's function due to the boundary potential, in terms of this \textit{local} change, 
\begin{equation}\label{eq:deltaG}
    \Delta G_{\rm tot}^{S^z_I,\sigma}(\omega) = \Delta G_{00}^{S^z_I,\sigma}(\omega) \times \left [ 1- \frac{\partial \Delta(\omega)}{\partial \omega} \right ] \;.
\end{equation}
From this we obtain the total change in the bath DoS, $\Delta\rho_{\rm tot}^{S^z_I,\sigma}(\omega)$, and hence the impurity contribution to the partition function from Eq.~\ref{eq:zfunc}. The impurity magnetization follows from Eq.~\ref{eq:mag}.

The full analytic solution of the generic Ising impurity model presented above yields some immediate insights, discussed below.


\subsection{$B_0=0$ case}
Keeping $B_I$ finite but setting $B_0=0$ results in an impurity response \textit{identical} to that of a free spin, independent of the specific bath used. This is because in this limit $\epsilon_{\textit{eff}}^{\uparrow\uparrow}=\epsilon_{\textit{eff}}^{\downarrow\downarrow}$ and $\epsilon_{\textit{eff}}^{\uparrow\downarrow}=\epsilon_{\textit{eff}}^{\downarrow\uparrow}$, meaning that $\prod_{\sigma}Z^{S^z_I,\sigma}_B$ is independent of $S^z_I$ in the calculation of the impurity spin-resolved partition functions 
$Z^{S^z_I}$. From Eq.~\ref{eq:mag}, these factors cancel out in the magnetization calculation and we obtain $\langle \hat{S}^z_I\rangle = -\tfrac{1}{2}\tanh(B_I/2T)$, as for an isolated free spin. Corrections to this result for small finite $B_0$ are expected to be small.


\subsection{Wide flat band limit}
Quantum impurity problems are often studied \cite{Hewson} using the flat band DoS Eq.~\ref{eq:dos_flb}, in the wide bandwidth limit $D\to \infty$. In practice, this applies for finite bandwidths which are the largest energy scale in the problem, $D\gg J^z, B_I, B_0, T$. In this case, the real part of the bath Green's function $G_{00}^0(\omega)$ vanishes, and so the auxiliary hybridization function takes the form  $\Delta(\omega)=\omega-i/(\pi\rho_0)$. From this, we see immediately that $\partial \Delta(\omega)/\partial \omega =1$ and hence $\Delta G_{\rm tot}^{S^z_I,\sigma}(\omega)=0$ in Eq.~\ref{eq:deltaG}. In turn, $Z^{S^z_I,\sigma}_B$ is independent of $S^z_I$ and $\sigma$ from Eq.~\ref{eq:zfunc}, and the isolated free spin result for the impurity magnetization $\langle \hat{S}^z_I\rangle = -\tfrac{1}{2}\tanh(B_I/2T)$ is again found from Eq.~\ref{eq:mag}.

\vspace{-0.5cm}
\subsection{Nanowire example}
For $B_0 \ne 0$ and arbitrary structured baths, we find a correction to the free spin result. However, in all cases considered, the deviation is rather modest. To illustrate this, we take the nanowire (1d chain) bath as a concrete example. The explicit form of Eq.~\ref{eq:Hbath_mod} then reads,
\begin{equation}\label{eq:Hbath_mod_1d}
    \hat{H}_{\textit{bath}}^{S^z_I,\sigma}= \epsilon_{\textit{eff}}^{S^z_I,\sigma} \left( \hat{n}^{\phantom{\dagger}}_{0\sigma}\right ) + t\sum_{j=0}^{\infty} \left (\hat{c}_{j\sigma}^{\dagger} \hat{c}_{j+1\sigma}^{\phantom{\dagger}} + \hat{c}_{j+1\sigma}^{\dagger} \hat{c}_{j\sigma}^{\phantom{\dagger}}\right )
    \;.
\end{equation}
The corresponding free bath Green's function at the impurity location is $tG_{00}^0(\omega)=\omega/(2t) - i\sqrt{1-(\omega/2t)^2} \equiv e^{-i\phi}$ with $\cos{\phi}=\omega/2t$. The local DoS seen by the impurity follows from $\rho_{\textit{nw}}(\omega)=-\tfrac{1}{\pi}{\rm Im}~G_{00}^0(\omega)$, and is given by Eq.~\ref{eq:dos_nw} with $D=2t$ and $\rho_0=1/\pi t$. The change in the local Green's function due to the boundary potential is still given by Eq.~\ref{eq:dyson}. 
Following Ref.~\cite{mitchell2011real}, we can express the change in the real-space Green's function for site $j \ge 0$ due to the boundary potential as $\Delta G_{jj}^{S^z_I,\sigma}(\omega)=\Delta G_{00}^{S^z_I,\sigma}(\omega)\times [t G_{00}^0(\omega)]^{2j}$. We note that the latter factors, $[t G_{00}^0(\omega)]^{2j}\equiv e^{-2ij \phi}$ are simply Chebyshev polynomials. Resumming the infinite series for all sites $j$, we obtain explicitly,
\begin{equation}
    \Delta G_{\textit{tot}}^{S^z_I,\sigma}(\omega)=\frac{\Delta G_{00}^{S^z_I,\sigma}(\omega)}{1-[t G_{00}^0(\omega)]^{2}} \;,
\end{equation}
which agrees with the general calculation from Eq.~\ref{eq:deltaG}. 

Using these expressions, we find that the correction to the free-spin magnetization is small except when the impurity-bath coupling $J^z \gtrsim D$ is very large.

\vspace{-0.5cm}
\subsection{Mean-field approximation}\vspace{-0.5cm}
Further physical insight is gained from a mean-field approximation for the Ising impurity magnetization problem.

The mean-field Hamiltonian for the Ising impurity model is obtained by decoupling the interaction term, assuming that impurity and bath spin fluctuations around their average values are small. We can then write,
\begin{equation}
	\label{Eq:MFT_Ham}
	\hat{H}_{MFT} = \hat{H}_{bath} + B_I^{\textit{eff}}\hat{S}_I^z + B_0^{\textit{eff}} \hat{S}_0^z \;,
\end{equation}
where the effective fields are $B_I^{\textit{eff}}=B_I + J^z \expval{\hat{S}_0^z}$ and $B_0^{\textit{eff}}=B_0 + {J^z}{\expval{\hat{S}_I^z}}$. 
Since the impurity and bath only talk to each other through their average magnetizations in mean-field theory, we can treat them as independent systems subject to the self-consistency condition,
\begin{align}
	\label{Eq:MFT_Imp_Mag}
	\expval{\hat{S}_I^z} &= -\frac{1}{2}\tanh{\left(\frac{B_I + {J^z}\expval{\hat{S}_0^z}}{{2}{T}}\right)} \;, \\
	\label{Eq:MFT_Bath_Mag}
	\expval{\hat{S}_0^z} &= \frac{1}{2}\left[\expval{\hat{n}_{0\uparrow}} - \expval{\hat{n}_{0\downarrow}}\right] \;.
\end{align}
Here $\expval{\hat{S}_0^z}$ depends on  $\expval{\hat{S}_I^z}$ through the effective boundary field acting on the bath, $B_0^{\textit{eff}}$.

We again develop a Green's function approach to solve the equations, so that we can use any bath DoS of our choosing. For a given free bath Green's function $G_{00}^0(\omega)$, we can incorporate the effect of the boundary field through the Dyson equation,
\begin{equation}
	\label{Eq:MFT_Dyson}
	\left[G_{00}^{\sigma}(\omega) \right]^{-1} = \left[G_{00}^0(\omega)\right]^{-1} - \sigma B_0^{\textit{eff}} \;,
\end{equation}
where $\sigma=\pm \tfrac{1}{2}$ for bath conduction electrons with spins $\uparrow / \downarrow$. The local spin-resolved occupancy then follows as
\begin{equation}
	\label{MFT:Average_Occup_No}
	\expval{\hat{n}_{0\sigma}} = -\frac{1}{\pi}\int_{-\infty}^{+\infty} d \omega {f(\omega)}~{\rm Im}\left [G_{00}^{\sigma}(\omega) \right ] \;,
\end{equation}
with $f(\omega)$ the Fermi function. 

Although a simple approximation, we find by comparison to the exact solution that in practice the mean-field result is extremely accurate, especially at larger $B_I$ and $T$. In particular, the form of Eq.~\ref{Eq:MFT_Imp_Mag} unveils the rather elegant physical interpretation that the Ising impurity magnetization is essentially that of a free spin, but with a renormalized effective field $B_I \to B_I^{\textit{eff}}$. Indeed, in most reasonable cases of interest we find the renormalization to be small.

\vspace{-0.8cm}
\bibliography{bibo}

\end{document}